\documentclass[12pt,preprint]{aastex} 		% single spaced
\usepackage{natbib}
\def\asec{$^{\prime\prime}$}
\def\amin{$^{\prime}$}

\def\Lsun{~L$_{\odot}$}
\def\deg{$^{\circ}$}

\begin{document}

\title{Epsilon Eridani's Planetary Debris Disk: \\
Structure and Dynamics based on \\
Spitzer and CSO Observations}

%\slugcomment{revised and resubmitted September 5, 2008} 
\shortauthors{Backman et al. } 
\shorttitle{Epsilon Eridani Debris Disk}

\author{D. Backman}
\affil{SOFIA \& SETI Institute}
\author{M. Marengo}
\affil{Harvard-Smithsonian Center for Astrophysics}
\author{K. Stapelfeldt}
\affil{Jet Propulsion Laboratory, California Institute of Technology}
\author{K. Su}
\affil{Steward Observatory, University of Arizona}
\author{D. Wilner}
\affil{Harvard-Smithsonian Center for Astrophysics}
\author{C. D. Dowell}
\affil{Jet Propulsion Laboratory, California Institute of Technology}
\author{D. Watson}
\affil{University of Rochester}
\author{J. Stansberry, G. Rieke}
\affil{Steward Observatory, University of Arizona}
\author{T. Megeath\altaffilmark{1}, G. Fazio}
\affil{Harvard-Smithsonian Center for Astrophysics}
\author{M. Werner}
\affil{Jet Propulsion Laboratory, California Institute of Technology}

\altaffiltext{1}{Now at the University of Toledo (Ohio)}

\begin{abstract}

{\it Spitzer} and Caltech Submillimeter Observatory (CSO) images
and spectrophotometry of $\epsilon$ Eridani at wavelengths from
3.5 to 350 $\mu$m reveal new details of its bright debris disk.
The 350 $\mu$m map confirms the presence of a ring at
$r =$ 11--28\asec~(35--90 AU) observed previously at
longer sub-mm wavelengths.  The {\it Spitzer}
mid- and far-IR images do not show the ring, but rather
a featureless disk extending from within a few arcsec of the
star across the ring to $r \sim$ 34\asec~(110 AU). 

The spectral energy distribution (SED) of the debris system
implies a conplex structure. A model constrained by the surface
brightness profiles and the SED indicates that the sub-mm ring
emission is primarily from large ($a \sim$ 135 $\mu$m) grains,
with smaller ($a \sim$ 15 $\mu$m) grains also present in and
beyond the ring.  The {\it Spitzer} IRS and MIPS SED-mode
spectrophotometry data clearly show the presence of spatially
compact excess emission at $\lambda \ga 15~\mu$m that requires
the presence of two additional narrow belts of dust
within the sub-mm ring's central void. The innermost belt at
$r \sim$ 3 AU is composed of silicate dust.

A simple dynamical model suggests that dust produced collisionally
by a population of about 11 M$_\oplus$ of planetesimals in the
sub-mm ring could be the source of the emission from both in
and beyond the sub-mm ring.  Maintaining the inner belts and the
inner edge to the sub-mm ring may require the presence of three
planets in this system including the candidate radial velocity
object.

\end{abstract}

\keywords{
infrared: stars -- circumstellar matter -- planetary systems -- 
stars: individual (epsilon Eridani)}

\section{Introduction}
\label{sec:intro}

The very nearby (3.22 pc; HIP 16537) K2 V star $\epsilon$ Eridani
(HR 1084, HD 22049, GJ 144) was found by IRAS to have a 
bright semi-resolved far-infrared excess due to circumstellar dust
(Gillett 1984; Aumann 1985).  The system shows $\geq 1$ Jy
of excess at 60--200 $\mu$m (IRAS 03305-0937; Walker \& Heinrichsen
2000 and references therein), and a moderate-significance excess at
25 $\mu$m.

VLTI angular diameter measurements plus evolutionary models
imply an age for $\epsilon$ Eri of 850 Myr (Di Folco et al.\ 2004),
consistent with previous determinations based on Ca II activity
(Henry et al.\ 1996), Li abundances (Song et al.\ 2000),
and gyrochronology (Barnes 2007).
For $\epsilon$ Eri, as well
as the other IRAS-discovered prototype debris disk systems
$\alpha$ Lyr (Vega), $\alpha$ PsA (Fomalhaut), and $\beta$ Pic,
(Gillett 1984; 1986) later dubbed the ''Fab Four'', estimated
time scales for dust destruction by Poynting-Robertson (P-R) drag and
mutual collisions are significantly less than the respective system ages.
Thus, the observed dust is not primordial but must
be ''second-generation'' material released relatively
recently, by collisions or other activity, from larger parent bodies.
These debris disks are therefore understood to
represent late stages, or remnants, of the planet formation
process.  Many more debris disks have been discovered since the
prototypes (see reviews by Backman \& Paresce 1993 (Protostars
and Planets III); Lagrange et al.\ 2000 (Protostars and Planets
IV); Meyer et al.\ 2007 (Protostars and Planets V)).
$\epsilon$ Eri is especially interesting for its
proximity, its age of nearly 1 Gyr, and its K main sequence spectral
type for which relatively few debris disks are presently known
(Trilling et al.\ 2008).

The $\epsilon$ Eri system was mapped at 850 and 450 $\mu$m
by Greaves et al.\ (1998; 2005), revealing a nearly circular ring
extending more than 25\asec~(80 AU) from the star, with peak
surface brightness at about 18\asec~($\sim$ 60 AU) and a central
cavity with radius about 10\asec~($\sim$ 30 AU).
Sch\"{u}tz et al. (2004) detected the ring at 1.2 mm and confirmed
its spatial extent.
Marengo et al.\ (2006)
made a deep search at wavelengths of 3.5 to 7.9 $\mu$m
for planetary-mass companions between 14\asec~(45 AU) and
2.9\amin~(560 AU) from the star with {\it Spitzer}'s IRAC camera,
and also set upper limits to extended emission.
Proffitt et al.\ (2004) used HST/STIS observations to determine an
upper limit to optical scattered light from the region of the
sub-mm ring.
Di Folco et al.\ (2007) set an upper limit
on near-IR (K' band) scattered light brightness
within 1\asec~(3 AU) of $\epsilon$ Eri
via CHARA interferometric measurements.

Dent et al.\ (2000), Li, Lunine \& Bendo (2003),
and Sheret, Dent \& Wyatt (2004) calculated disk models
constrained by the surface brightness distribution in
the sub-mm images and pre-{\it Spitzer} SED data,
examining alternatives of porous or solid
grain structures and silicate or silicate-plus-ice compositions.
Bright spots in the sub-mm ring have been interpreted as
dust density enhancements caused by a planet orbiting in or
near the ring (Liou \& Zook 1999; Ozernoy et al.\ 2000;
Quillen \& Thorndike 2002; Deller \& Maddison 2004;
Greaves et al.\ 2005).

Tentative detection of a Jovian-mass planet
orbiting $\epsilon$ Eri with
a semi-major axis of 3.4 AU, well inside the sub-mm ring,
has been made via radial velocity measurements
(Campbell et al.\ 1988; Hatzes et al.\ 2000), possibly
confirmed astrometrically (Benedict et al.\ 2006).
Das \& Backman (1992) suggested that
such a planet might influence the location of the warm
dust detected by IRAS.
Detailed models of the gravitational effect of a planet 
close to the star on possible nearby dust distributions
were calculated by Moran, Kuchner \& Holman (2004).

The present paper is part of the {\it Spitzer} all-instrument GTO
''Fab Four'' program making comprehensive studies of the prototype
planetary debris disks, to put our solar system into evolutionary
context and to search for signs of planets
(Fomalhaut:  Stapelfeldt et al.\ 2004;
Vega: Su et al.\ 2005;
$\beta$ Pic:  Chen et al.\ 2007).
{\it Spitzer} obtained the first far-IR images
of $\epsilon$ Eri, which yield broadband
photometry and disk surface brightness
profiles as well as high-resolution comparisons with prior
sub-mm maps.  Those {\it Spitzer} images are combined here
with medium-resolution IR spectroscopy plus a 350 $\mu$m map
from the CSO SHARC II camera.
Together, these data provide a unique opportunity
to study the nearest known exo-Kuiper belt and exo-zodiacal
dust system in detail.

Section 2 of this paper describes the {\it Spitzer} and CSO
observations of $\epsilon$ Eri and the data reduction.
Section 3 addresses detailed analyses of the data including
first-order corrections of image and spectrophotometric data
plus determination of the photosphere SED and IR excess.
Section 4 describes derivation of
a model of grain spatial and size distributions
fitted to the disk's multi-$\lambda$ surface brightness
distributions and over-all SED.
Section 5 discusses implications of the model results,
compares them with other observations and SED-based
models of the $\epsilon$ Eri disk,
and presents results of a simple dynamical model
calculation connecting $\epsilon$ Eri's disk properties
with the history of our solar system's Kuiper belt.

\section{Observations and Data Reduction} 
\label{sec:obs+redux}

{\it Spitzer} observations of $\epsilon$
Eri were taken in Guaranteed Time Observing Program 90 (P.I.
Michael Werner). 
Data obtained with various instruments and facilities were used as follows:
(1) {\it Spitzer} IRAC image photometry measures the photospheric SED
from 3.6 to 7.9 $\mu$m, wavelengths that show no detectable contribution
from the debris disk.
(2) {\it Spitzer} MIPS data at 24, 70, and 160 $\mu$m
provide mid- and far-IR images of the disk.
(3) {\it Spitzer} MIPS SED-mode data give source flux
densities from 55 to 90 $\mu$m with low spectral resolution,
plus source size in one dimension versus wavelength.
(4) {\it Spitzer} IRS spectra link the mid-IR
broadband photometric points,
determine the shortest wavelength of excess (highest
abundant dust temperature, minimum dust orbit radius),
and allow a search for mineralogical features.
(5) The CSO 350 $\mu$m map can be compared
with the {\it Spitzer} far-IR images as well as
with prior 450 and 850 $\mu$m maps.
Surface brightness profiles from the 24, 70, 160 and 350 $\mu$m
data together strongly constrain the disk model.
(6) The {\it Spitzer} MIPS images and CSO map are also integrated
to yield broadband photometric points for the entire disk,
supplemented by IRAS catalog photometry using re-calculated color
corrections, plus previously published ISO photometry.

\subsection{Spitzer IRAC data}
\label{sec:irac_data}

Images at 3.6 to 7.9 $\mu$m of $\epsilon$ Eri and its surroundings
were made with {\it Spitzer}'s InfraRed Array Camera (IRAC;
Fazio et al.\ 2004) in 2004 January and February.
Those observations were designed to look 
for wide-separation low-mass companions, but also provided accurate
photometry of the star via Point Spread Function (PSF) fitting
of non-saturated portions of $\epsilon$ Eri and standard star images.
Details of the observations and data reduction procedures
are described by Marengo et al.\ (2006).
The IRAC data were reprocessed for the present paper to take
advantage of an updated instrument calibration plus
availability of additional reference star observations
to generate a new PSF.

Each of the two IRAC observations consisted of a 36-position dither
of the source on the array, for a total of 2 x 3380 seconds integration time.
The new PSF reference stars were $\epsilon$ Indi (K4.5 V), Vega, Fomalhaut
and Sirius.  (The debris disks around Vega and Fomalhaut
are not detected at IRAC wavelengths.)
The total field of view had a width of 5.78\amin~due to the dither,
slightly larger than the IRAC frame width of 5.21\amin.  Data 
reduction and calibration were done using the {\it Spitzer}
Science Center (SSC) 
pipeline version S15 combined with the post-BCD IRACproc software
(Schuster, Marengo \& Patten 2006).

\subsection{Spitzer MIPS data}
\subsubsection{MIPS 24, 70 and 160 $\mu$m Images}

Observations of $\epsilon$ Eri were made using all three
imaging channels of the Multiband Imaging 
Photometer for {\it Spitzer} (MIPS; Rieke et al.\ 2004).
Images at 24 $\mu$m were made on 2004 January 29 using two
different observing strategies.  (1) The standard 
24 $\mu$m MIPS photometry dither pattern was executed at four cluster target 
positions in a square pattern 8.29\asec~(3.25 pixels) wide.
This procedure
yielded an exposure time of 4$\times$(16$\times$3) seconds
for the region within  
2.5\amin~of the star, with a high degree of sub-pixel sampling to 
facilitate PSF subtraction.  Observations of
the K4 III star HD 217382 made on 2003 November 19 with the identical
dither/cluster 
target pattern were used as a PSF reference.
(2) To probe a wider field of view for possible companions, another
$\epsilon$ Eri 24 $\mu$m dataset was taken using a $3\times3$ raster map.  
The resulting mosaic covers a 0.25\deg$\times$0.25\deg~region 
centered on the star to an exposure depth of 2$\times$(16$\times$3) seconds.

Two types of MIPS 70 $\mu$m image observations were also taken
on 2004 January 29.  (1) With the default pixel scale
($9.96$\asec/pixel), eight cycles of the large-field photometry
dither pattern were used, providing a total exposure 
time of 151 seconds per pixel.
(2) With the fine pixel scale ($5.2$\asec/pixel), one cycle of the small 
field photometry dither pattern was made at each of 12 cluster target 
positions, arranged in a rectangular grid with 16.22\asec~(3.25-pixel)
spacing, for an effective exposure time of 12$\times$(10$\times$3)
seconds per pixel.
On 2005 April 4 a deeper fine scale image was obtained with an
effective exposure time of 12$\times$(10$\times$10) seconds per pixel.

Two MIPS 160 $\mu$m image datasets were obtained.
(1) The first set was taken on 2004 January 29, using 2 cycles of
small-field photometry at each of 9 cluster target positions spaced
on a 36\asec~rectangular grid, and 3-second exposures, for an exposure
time of 9$\times$27 seconds per pixel.  (2) Another set
was obtained on 2006 Feb 19 using the same observing
strategy except that 10-second exposures were used,
for an exposure time of 9$\times$90 seconds.

The MIPS 160 $\mu$m camera suffers from a spectral 
leak by which stray light in the wavelength range 1-1.6 $\mu$m
produces a false image partially overlapping the actual
160 $\mu$m source image.
This ghost image is several times brighter than the real disk
emission in the case of $\epsilon$ Eri, and must be subtracted
accurately to enable study of the circumstellar emission.
The spectral leak is field-dependent, and therefore can only be 
properly subtracted using a reference star that was observed with the
same dither  
parameters as the science target.  Pre-launch observation planning did
not anticipate this requirement, so no appropriate reference star
observation was available initially to leak-subtract the $\epsilon$ Eri
dataset.

Good leak subtraction was achieved using 160 $\mu$m
images taken on 2004 August 6
of the debris disk star $\tau$ Ceti (type G8)
with the same dither pattern as for $\epsilon$ Eri.
Only the spectral leak, and not the photosphere or disk excess,
is detected at 160 $\mu$m from $\tau$ Ceti
(Stapelfeldt et al., in preparation).
The subtraction procedure involved empirically determining the maximum
normalization factor for the leak reference source such that its subtraction
from the science target did not produce noticeable residuals below the
background level.  The subtractions were performed using mosaics
in detector coordinates of all the dithered observations.
Observations of Achernar ($\alpha$ Eri) were
also obtained for $\epsilon$ Eri leak subtraction reference,
on 2005 August 30.  The Achernar data did not support a good
leak subtraction for $\epsilon$ Eri,
probably because of the large color difference between
the two stars (type B3 versus K2).  

The MIPS instrument team's Data Analysis Tool (DAT; Gordon 
et al.\ 2005) was used for basic reduction (dark subtraction, flat
fielding/illumination correction) of the SSC S11 data products.
The 24 $\mu$m images were processed 
to remove a vertical ''jailbar'' pattern along the detector columns
that appears during observations of bright sources.  The known transient 
behaviors associated with the MIPS 70 $\mu$m array were removed
from the default-scale data by median column 
subtraction and time-filtering the images with an excluded region of 15
pixels ($\sim$ 150\asec) centered on the source
(for details see Gordon et al.\ 2007).  For the 70 $\mu$m
fine-scale data, off-source exposures were subtracted from
preceding on-source exposures to remove the time-dependent
background column offsets.
Flux calibration factors of
0.0454 MJy sr$^{-1}$ (24 $\mu$m),
702 MJy sr$^{-1}$ (70 $\mu$m default-scale),
2894 MJy sr$^{-1}$ (70 $\mu$m fine-scale), and
44.7 MJy sr$^{-1}$ (160 $\mu$m)
per MIPS raw data unit
were applied to the MIPS image data
(Engelbracht et al.\ 2007; Gordon et al.\ 2007;
Stansberry et al.\ 2007).
The processed data were then combined using
World Coordinate System (WCS) information to produce final mosaics
with pixels half the size of the physical pixel scale.

\subsubsection{MIPS SED-mode Data}

MIPS Spectral Energy Distribution (SED-mode) data
were obtained on 2005 September 4 using 10 cycles of
10 second exposures with a +3\amin~chopper throw. The total
integration time was 629 seconds.  Basic reduction of the data to produce
calibrated versions of the individual exposures was done using version
3.06 of the MIPS DAT code.  MIPS SED-mode observations provide
chopped pairs of images.  The off-source images were pair-wise
subtracted from the on-source images, and the results mosaicked
to produce an image of the 55--90 $\mu$m spectrum of $\epsilon$ Eri.
The mosaic was generated using 4.9\asec~pixels, yielding an image
65 pixels across in the spectral dimension by 44 pixels in the spatial
dimension.  This image was then boxcar-smoothed over 5 adjacent rows,
resulting in spectral resolution $R =$ 15--20.

\subsection{Spitzer IRS data}

Mid-infrared spectra of $\epsilon$ Eri were obtained in
2004 February, 2004 August, and 2005 January with the
{\it Spitzer} Infrared Spectrograph (IRS; Houck et al.\ 2004).
Using the low-spectral resolution
($R \sim$ 60--120)
spectrographs SL1 (7.5--14 $\mu$m), LL2 (14--21 $\mu$m)
and LL1 (20--38 $\mu$m), observations were made with
five slit positions separated
by half a slit width (1.8\asec) in SL1 and
ten slit positions separated by half a slit width
(4.8\asec) in LL2 and LL1.
The star was centered in the scan pattern in each case.
Staring-mode observations were also obtained using the
low-resolution SL2 (5.3--7.5 $\mu$m) and high resolution
($R \sim$ 600) SH (10--20 $\mu$m) and
LH (19--37 $\mu$m) modules.  Those observations were at two nod
positions, with the star displaced from the center of the slit
by a third of the slit length.
Total integration times per slit position or nod were 84 seconds
in the SL1 spectral map, 70 seconds each in the LL2 and LL1 spectral
maps, and 54 seconds, 72 seconds, and 144 seconds respectively
for the SL2, SH and LH observations.
Each set of observations was preceded by a high-accuracy
pointing peak-up observation of the neighboring star SAO 130582
using the IRS Blue (15 $\mu$m) camera.

Data reduction began with products of the {\it Spitzer}
Science Center's S11 IRS data pipeline.  Further processing
involved removal, by interpolation in the spectral direction, of
permanently bad and ''rogue'' pixels identified in the IRS
dark-current data for all campaigns up to and including the
one in which each $\epsilon$ Eri spectrum was taken.
The IRS instrument team's SMART software was then used to extract
1-D spectra from the 2-D long-slit (SL, LL) or echelle (SH, LH) images.
Sky emission in the SL and LL observations was measured by a linear fit
to the signals at points along the slit farther than 40\asec~from the star,
and subtracted as part of the extraction.  The
$\epsilon$ Eri system appeared to be spatially unresolved
in the SL observations, so an extraction window was used that was
fitted to the instrumental point-spread function at the longest
wavelengths of each low-resolution module, 4 pixels wide
in each case.  A full-slit extraction was performed for
the high-spectral-resolution (SH and LH) observations.
The sky emission measured in the
low-spectral-resolution observations was small enough compared
with the flux received from $\epsilon$ Eri and its disk
that sky subtraction in the SH and LH observations
could be neglected.

Reference SL and LL observations of $\alpha$ Lac (A1 V)
plus SH and LH observations of $\xi$ Dra (K1 III)
were obtained in the same manner as the $\epsilon$ Eri observations.
The spectra of $\epsilon$ Eri were calibrated via division by
spectra of the respective reference star taken at the corresponding
nod position, multiplication by a synthetic template spectrum of the
reference star (Cohen 2004, private communication), and tben nod-averaging.
Reduced spectra from each spectrograph module were then
combined to produce complete 5.3--38 $\mu$m spectra.
Small spectrograph module-to-module differences
at wavelengths of overlap were assumed to be due to minor
telescope pointing-induced throughput losses, and removed by
scaling up the spectra from smaller-slit modules to
match those from the larger-slit modules in the overlap range.

Linearity and saturation in the IRS detector arrays is corrected
in the data pipeline, but $\epsilon$ Eri and its disk are bright
enough that linearity and saturation effects in the
low-spectral-resolution modules are
large and the corrections uncertain.  The main impact
of these effects is that the LL spectra
had significantly lower effective signal-to-noise ratio and
spectrophotometric accuracy than the SH and LH observations
due to residual inaccuracy of the LL saturation correction.
The final spectrum used in this paper was assembled from
SL, SH and LH data; LL data will not be discussed further.
The uncertainty in flux density is 5\% below
about 10 Jy, increasing to about 10\% for flux
densities as high as 20 Jy.

\subsection{CSO Data}
\label{sec:cso_data}

Observations of $\epsilon$ Eri at 350 $\mu$m
were made during parts of three nights in
2003 January and two nights in 2005 December using the
SHARC II 12$\times$32 bolometer array (Dowell et al.\ 2003)
at the Caltech Submillimeter Observatory (CSO).
The weather conditions were excellent, with 225 GHz (1.3 mm)
zenith opacity ranging from 0.028 to 0.048.
The total integration time on the source was 11 hours.
Data were obtained by rapid scanning (without chopping)
in Lissajous patterns with periods of 14--20 seconds
and amplitudes of $\pm$15\arcsec~to $\pm$40\asec~in each
of two axes.

Each SHARC II detector spans 4.6\asec, and the
diffraction-limited beam size is 9\asec~(FWHM), with
a PSF determined by coadding images from
a variety of calibration sources
interspersed with the $\epsilon$ Eri observations.
Data reduction used the {\it sharcsolve} software package that
iteratively decomposes fixed source emission, time variable
atmospheric emission, and detector offsets.  Best results for
subtraction of the temporally and spatially variable telluric
emission were achieved by assuming that the average emission
outside a 35\asec~radius from the star was zero, at the cost
of filtering out possible more widely extended emission.
Calibration was based
on the integrated map signals within a 35\asec~radius, following
correction for atmospheric absorption, from sub-mm
standards OH231.8 (18 Jy), CRL618 (19 Jy), o Ceti (2.3 Jy),
Ceres (13 Jy in 2003 January), and Juno (8.0 Jy in 2005 December).
The uncertainly in the absolute flux scale is estimated to
be 25\%, and the RMS noise level in the reduced 350 $\mu$m
image is $\sim 3$ mJy per smoothed beam.

\section{Results}
\label{sec:results}

\subsection{Images and Spatial Structure}

\subsubsection{MIPS Far-IR Images}

The top row of Figure 1 displays the MIPS 24, 70 and 160 $\mu$m
images after standard {\it Spitzer} pipeline processing.
The bottom row of Figure 1 shows the output from a further level of
processing.  Specifically: (1) Figure 1d is the result of subtracting
a scaled PSF from the 24 $\mu$m image in 1a, (2) Figure 1e shows a
70 $\mu$m fine pixel-scale image for comparison with the default
pixel-scale image in 1b, and (3) Figure 1f is the result of compensating
for the 160 $\mu$m filter's short-wavelength leak apparent in 1c.
Radial profiles of the processed images are presented and discussed
in Section 4.2.1.

The 24 $\mu$m PSF-subtracted data show an image core
size of 5.6\asec$\times$5.4\asec, consistent with
the values of 5.52\asec$\times$5.48\asec~for an unresolved
blue (photospheric) source determined by Su et al.\ (2008).
The stripes in the processed data are PSF subtraction
residuals, mainly noise arising from different sampling
of the source and PSF reference by the pixel grid.
Examination of the 24 $\mu$m PSF wings shows evidence
for slight excess surface brightness
at the position of the first Airy dark ring ($r \sim$ 7\asec)
but with only $\sim1\sigma$ significance.

The far-IR source is resolved at 70 and 160 $\mu$m.
At 70 $\mu$m (fine pixel-scale image)
the observed source has a compact core
with FWHM of 24.2\asec$\times$22.3\asec~elongated
approximately north-south, superimposed on a relatively
flat pedestal of emission roughly 1\amin~in diameter.
The core 70 $\mu$m emission lies mostly {\it interior} to
the sub-mm ring, and is clearly extended relative to
the size of unresolved point sources, which is
15.3\asec$\times$14.4\asec~(Su et al.\ 2008).
The leak-subtracted 160 $\mu$m source can be approximated
by Gaussian profiles with FWHM 62\asec$\times$52\asec,
also elongated north-south.
This distribution is also extended
relative to a point source (39\asec~diffraction FWHM).
The stellar point source contributes about
10\% of the total system flux density at 70 $\mu$m
and 5\% at 160 $\mu$m.

The 70 and 160 $\mu$m images contain a substantial surprise.
Although the far-IR source FWHM is about the size of the sub-mm ring,
the far-IR images do not show the ring-with-deep-hole
aspect observed at longer wavelengths.
Instead, the far-IR surface brightness distribution is continuous
across the central region and the sub-mm ring,
extending well beyond the ring,
with no significant enhancement at the position of the ring.
The far-IR source is close to circularly symmetric,
with possible slight extension north-south as in the
350 and 450 $\mu$m source maps, but unlike the 850 $\mu$m maps
that are extended east-west or northeast-southwest
(Greaves et al.\ 2005).

To test for far-IR detection
of surface brightness features seen in the sub-mm ring,
an azimuthal average of the 70 $\mu$m fine-scale image
was subtracted from the image itself.
Figure 2a shows the result of this analysis.
The subtracted image is smooth to $\la$ 20\% of
the surface brightness in the ring, smaller than the
amplitude of features in prior sub-mm maps.
The 70 $\mu$m image therefore does not show
the azimuthal surface brightness variations
observed at longer wavelengths. A similar result, albeit with
lower signal-to-noise ratio, is obtained with the
160 $\mu$m image (Figure 2b).

\subsubsection{MIPS SED-mode Source Size}

Figure 3a shows the MIPS SED-mode source FWHM
(along the slit) versus wavelength from Gaussian fits
to the spatial profiles of the signal at each row of the mosaic image.
The FWHM are 21, 23.5, and 38\asec~at 55, 70 and 90 $\mu$m,
respectively, consistent with the 70 and 160 $\mu$m image source sizes.
Analysis of spectrophotometric data from the
MIPS-SED instrument, presented in Figure 3b,
is described in Sections 3.3.1 and 4.3.2.
The small-scale variations in FWHM and in
flux density versus wavelength relative to simple
monotonic trends, as well as the steep increase in FWHM
longward of 87 $\mu$m, are not real but represent
systematic uncertainties in the reduction procedures,
which dominate photometric uncertainties.

\subsubsection{CSO Sub-mm Map}

Figure 4 displays the CSO SHARC II 350 $\mu$m map.
The sub-mm ring that has been observed previously at
450 and 850 $\mu$m is apparent here, with approximately the same spatial scale.
The 350 $\mu$m image has flux consistent with zero excess in
the central beam (FWHM $=$ 29 AU)
after subtraction of a model 8 mJy point source corresponding
to the stellar photosphere emission, representing
2\% of the total flux in the map (Table 1).
The 350 $\mu$m ring may show north-south extension, as in
the {\it Spitzer} 70 and 160 $\mu$m images and the JCMT
450 $\mu$m map, but differing from the 850 $\mu$m morphology.

The average value and dispersion of the surface brightness around
the ring at 350 $\mu$m is S$_{\nu} = 0.15 \pm 0.04$ mJy per square arcsec.
The brightest 350 $\mu$m spot, in the north to north-east sector of
the ring, has S$_{\nu}$ $\sim$ 0.24 mJy per square arcsec,
about 2$\sigma$ above the average.
The location and shape of the 350 $\mu$m ring are confirmed by
the 2003 and 2005 data sets analyzed independently, but the
azimuthal substructure is not.

\subsection{Broadband Photometry}

\subsubsection{Stellar Photosphere}

Precise determination of $\epsilon$ Eri's photospheric SED was
necessary to measure the excess emission from the disk, especially
in the mid--IR where the system SED is only slightly above the stellar
continuum.
The star has m$_v$ = +3.74,  B-V = +0.88 (Mendoza et al.\ 1978)
and V-K of +2.11 (Koornneef 1983a).
Average main sequence B-V colors are +0.82 and
+0.92 for K0 and K2 temperature classes, respectively (Johnson 1966).
Average V-K colors are +2.00 and +2.25 
for K1 and K2, respectively (Koornneef 1983b).  These data
imply that $\epsilon$ Eri is a little bluer than the average K2 V
star, with interpolated temperature class between K1 and K2.
Flower (1996) gives T$_{eff}$ = 5090 K for B-V = 0.88.

Flux densities from $\epsilon$ Eri at 3.6, 4.5,
5.7 and 7.9 $\mu$m determined by PSF-fitting of {\it Spitzer}
IRAC data (Section 2.1) have ratios nicely consistent with expectations
for a photosphere of this star's temperature and luminosity.
The photosphere model SED used in this paper is a Kurucz
model for T = 5000 K (nearest match to 5090 K) and log g = 4.5,
normalized to the four IRAC flux densities
via least-squares minimization.  The RMS of the fit to those
points is 1.5\%.  The result is consistent with ground-based
photometry at 2.2 $\mu$m and shorter wavelengths
to better than 2\% precision.  Figure 5 displays the IRAC
data points and fitted photosphere in relation to other
broadband measurements discussed below.  Table 1
presents extrapolated photospheric flux densities which
can be regarded as having 2\% uncertainty.

\subsubsection{Photometry from Image Data}

Since the disk is not resolved at 24 $\mu$m, the total emission
from the disk at that wavelength was
obtained using both aperture photometry and PSF-fitting photometry
optimized for point sources; for details see Su et al.\ (2006).
A total of $2.04 \pm 0.04$ Jy was estimated for the star and disk
combined, which is 18\% above the expected photosphere.
No color correction was applied at 24 $\mu$m because the observed
SED slope in that wavelength region implies a MIPS filter correction
factor of $\sim$ 1.

The total flux density of the system at 70 and 160 $\mu$m was
integrated over a circular aperture with radius
64\asec~centered at the star.
The total disk emission in the 70 $\mu$m and 160 $\mu$m bands
is respectively 1.50 and 0.93 Jy after application of
color correction factors of 0.893 and 0.971
appropriate for a 55 K blackbody.

IRAS photometry of $\epsilon$ Eri spans an important range of
wavelengths, from the red end of the photospheric SED to beyond
the peak of the excess. 
The {\it Spitzer} IRS spectrum, MIPS images and MIPS SED-mode
data allow a refined
estimate of the continuum slope across the IRAS 25 and 60 $\mu$m bands
and hence more accurate color corrections, which were applied here.

Figure 5 and columns 1 \& 2 of Table 1 present the merged
system (star+disk) broadband SED
including {\it Spitzer} MIPS photometry from
image data, integrated flux density from the CSO 350 $\mu$m map,
newly color-corrected IRAS catalog flux densities,
ISO 200 $\mu$m flux density,
and previously published integrated flux densities
at 450 and 850 $\mu$m from Greaves et al.\ (2005).
The peak of the SED is between $\lambda$ $\sim$ 70 and 100 $\mu$m.

ISO photometry of $\epsilon$ Eri (Walker \& Heinrichsen 2000)
yielded a flux density of 1880 mJy at 200 $\mu$m (cf.\ predicted
photosphere value of 25 mJy), far above the 160 $\mu$m point.
The {\it Spitzer} 160 $\mu$m image, as well as the 450 and 850 $\mu$m
sub-mm maps, show significant background emission extended
beyond 40\asec~from the star that would have been included
in the large long-wavelength effective beam sizes of both ISO
(200 $\mu$m photometric aperture diameter 90\asec) and
IRAS (100 $\mu$m detector footprint $\sim$ 180\asec$\times$300\asec).
Numerous objects with near-IR colors of galaxies that would be
strong sources at $\lambda \ga$ 100 $\mu$m are detected in
{\it Spitzer} IRAC images of the $\epsilon$ Eri field
(Marengo et al.\ 2006).
The presence of this complex background emission probably
accounts for the high flux density measured by ISO;
the present observations indicate that much of the ISO
flux must come from beyond the sub-mm ring.
The IRAS 100 $\mu$m and ISO 200 $\mu$m photometry
(Figure 5) therefore provide upper limits
to the system flux densities at those wavelengths.
A less confused view of the source at 100--200 $\mu$m
should be possible eventually with deep Herschel PACS and SOFIA HAWC
observations.

\subsection{Spectrophotometric Data}
\label{sec:spectrophotometry}

\subsubsection{MIPS SED-mode}

With $\epsilon$ Eri expected to appear as an extended source along
the MIPS SED slit, simple aperture photometry measurements cannot be
made without admitting significant background noise.  To address
that problem, a Gaussian was fit to the spatial profile of the signal
along each row of the SED mosaic.  The integrated product of the Gaussian
width and height gives the source response at each wavelength.
The resulting $\epsilon$ Eri spectrum in instrumental units was converted
to physical units by reducing and extracting an archival spectrum of the
calibration star Canopus ($\alpha$ Car, G8 III) in the same way.
The Canopus spectrum
was normalized to give a flux density of 2.15 Jy at 70 $\mu$m
(the star's 3.11 Jy source flux density multiplied by a 69\% slit
throughput factor appropriate for a point source).
The result was then corrected for
the spectral response function by assuming Canopus has a Rayleigh-Jeans
spectral slope in the far-IR.  The flux normalization and spectral
response function derived from Canopus were then applied to the
$\epsilon$ Eri spectrophotometry.

Figure 3b presents the resulting MIPS SED-mode spectrum 
from 55 to 90 $\mu$m, giving the detected flux density transmitted
through the MIPS SED slit independent of assumptions about the true
spatial extent of the source.
The slope of the spectrum, declining to longer wavelengths,
is due to the intrinsic shape of the star+disk SED
combined with wavelength-dependent slit losses as the size
of the diffracted and extended emission region varies relative
to the 20\asec-wide MIPS SED entrance aperture.
Linking the MIPS SED-mode spectrophotometric data
to a disk model is described in Section 4.3.2.

\subsubsection{IRS}

{\it Spitzer} IRS spectrophotometric data for the disk plus stellar
photosphere are included in Figure 5.
The photosphere model described in Section 3.2.1
is overplotted for comparison.
A constant offset of 173 mJy was subtracted from the IRS
data, determined by making the average excess above the
photosphere model from 5 to 12 $\mu$m equal zero.
This is consistent with the observed lack of significant excess
emission in bradband photometry at 5.7, 7.9 and 12 $\mu$m (Table 1).
That offset represents only 1--3\% of the total source
flux in that wavelength range, well within the IRS calibration
uncertainty.
The resulting rescaled IRS mid-IR SED is in close
agreement with the MIPS 24 $\mu$m image photometry
as well as with the lower-precision IRAS 25 $\mu$m broadband point.

The Figure 5 panel inset shows the
spectrum from 10 to 30 $\mu$m after subtraction of the photosphere.
Significant departure of the
$\epsilon$ Eri SED from the photosphere begins at about 15 $\mu$m.
The mid-IR SED is remarkable in having a very steep rise from
15 to 20 $\mu$m, a ''plateau'' of almost constant flux density
from about 20 to 30 $\mu$m, and then the beginning of another steep
rise beyond 30 $\mu$m.

\section{Modeling}
\label{sec:modeling}

\subsection{General Considerations}
\label{sec:general}

The data presented above provide new information on the
$\epsilon$ Eri circumstellar dust spatial distribution, new
spectral information regarding warm dust components, and
separation of disk emission from confusing background sources.
A model of particle spatial distributions and physical properties
can now be produced that will allow improved understanding of dust
lifetimes and dynamics, which in turn allows inferences about
locations of planetary companions, and about the evolution of this
system for comparison with our solar system.

The modeling process involved: (1) derivation of
intrinsic 70, 160 and 350 $\mu$m 1-D radial surface brightness
profiles that, when convolved with the respective PSFs, match the
observed radial profiles; (2) calculation of a physical model
for the outer disk with specific grain spatial distributions,
size distributions and compositions that yields the intrinsic
70, 160, and 350 $\mu$m surface brightnesses inferred from the
first step; (3) derivation of another physical model for warm
material detected by IRAS and {\it Spitzer} IRS
but not resolved in the imaging data, then testing that the
predicted long-wavelength emission from the inner
material would be consistent with the 70, 160 and 350 $\mu$m
radial profiles; and finally, 4) checking that the over-all
physical model, including both resolved and unresolved material,
correctly predicts the observed spectral energy distribution of IR excess
from 15 to 850 $\mu$m and also the MIPS SED-mode narrow-slit
flux densities plus 1-D source sizes or limits.
As mentioned in Section 3.3.1, the MIPS SED fluxes need to
be corrected for slit losses, and that can only be done
by estimating the corrections iteratively using synthetic
images derived from the model (Section 4.3.2).

The 350 $\mu$m map shows a deficit of sub-mm emission
at $r \la 35$ AU relative to 35--90 AU, similar to previous sub-mm maps.
In contrast, the 24, 70 and 160 $\mu$m image profiles and photometry
plus mid-IR spectrophotometry and broadband photometry,
as well as the original IRAS detection, together
indicate the presence of significant amounts of warm material lying
at $r \la 35$ AU, within the central void in the sub-mm ring.

The surface brightness in the high-S/N 70 $\mu$m images
is continuous and monotonic out to
r $\ga$ 45 arcsec (140 AU), far outside the sub-mm ring.
However, trial PSF convolutions
of model intrinsic brightness distributions
revealed that the outer boundary of the 70 $\mu$m
emission is actually located at $r \sim$ 34\asec~(110 AU),
with apparently further extended emission representing the PSF wings.
In comparison, nearly all of the observed sub-mm surface
brightness is contained in the ring between about 35 and 90 AU
(11--28\asec) (Greaves et al.\ 2005; also, see
Sch\"{u}tz et al.\ 2004 regarding the source diameter at 1.2 mm).
The region outside the sub-mm ring, beyond $r = 28$\asec~(90 AU),
is here designated the ''halo''.

An outer limit to the 160 $\mu$m emission is more difficult to
determine due to background source confusion discussed above,
but the data and preliminary modeling indicate that the over-all
spatial distribution of material observed at 160 $\mu$m much more
closely resembles the 70 $\mu$m than the 350 $\mu$m emission.
Significant emission at 350 $\mu$m
does not extend beyond 28\asec~(90 AU).

\subsection{Outer Fully Resolved Disk}

\subsubsection{Radial Surface Brightness Profiles}

The 70 $\mu$m images have small ellipticity
and also are azimuthally smooth (Figure 2),
justifying uses of 1-D average radial profiles centered
on the star position for detailed comparison of model
surface brightness distributions to observations.
Based on inspection of the radial profiles and preliminary
modeling, fiducial radii of $r =$ 11, 28, and
34\asec~(35, 90, and 110 AU), representing transition
points in some of the surface brightness distributions,
were chosen as primary points for matching models to observations.

Intrinsic surface brightness profiles at 70, 160 and
350 $\mu$m were constructed by choosing surface brightness
values at the three fiducial radii and interpolating between those
points with simple power laws.
Production of model PSF-convolved radial profiles for
trial-and-error comparison with the
observations involved expansion of model 1-D radial profiles
into 2-D pseudo-images, convolution of each pseudo-image with
the respective wavelength's 2-D PSF derived for a point source
with $\epsilon$ Eri's temperature using STinyTim (Krist 2006),
and then collapsing the convolved model 2-D map back into a 1-D
profile.  After each trial, differences between predicted
and observed surface brightness profiles were used to estimate
new values at the fiducial radii for the next trial.

The results of modeling the 24, 70, 160 and 350 $\mu$m
surface brightness profiles are displayed in Figure 6.
The actual locations of changes in profile slopes near the
fiducial radii are uncertain to probably 10\% in radius, i.e.\ the
modeling process had approximately that much spatial resolution.
Note that these intrinsic surface brightness profile
determinations were aimed merely to match the 
observed profiles, without reference at this step
in the modeling process to physically motivated grain
properties, which are treated in the next section.

\subsubsection{Physical Grain Models}

The Heidelberg DDS code (Wolf \& Hillenbrand 2005)
was used to estimate grain properties
and surface densities that would yield the surface brightness values
determined in the preceding modeling step.
Grains were assumed to be solid (non-porous) and spherical.
DDS settings represented
$\epsilon$ Eri as a 5090 K blackbody with 0.33\Lsun.
Choice of grain composition, grain minimum and maximum sizes,
size distribution power law exponent, and mass surface densities
at the three fiducial radii allowed calculation of emitted surface
brightness for comparison
with the intrinsic 70, 160 and 350 $\mu$m surface brightness
values determined at those locations.
Integrated disk flux densities and grain masses
could then be calculated using
surface densities and grain properties interpolated
between the values at the fiducial radii.

The large difference in appearance of the sub-mm maps versus
the {\it Spitzer} 70 and 160 $\mu$m images led to a search
for model disk components such that:
(1) material having high sub-mm and low far-IR
emissivity would dominate in the sub-mm ring region,
coexisting with (2) material having low sub-mm and high far-IR
emissivity extending across the ring into the halo region.
Many combinations of grain properties, size distributions
and spatial distributions can match those constraints.
A simple model can be built of
components that separately fit the 70 \& 160 $\mu$m images
and the 350 $\mu$m map.

The final model for the resolved disk combines:
(1) large (radii $a =$ 100--200 $\mu$m) amorphous H$_2$O ice
grains located in the sub-mm ring, plus (2) small
($a =$ 6--23 $\mu$m) ''astronomical'' silicate grains having
a broader spatial distribution that includes the ring.
Experiments varying the value of the grain size
distribution exponent indicated that the conventional
collision equilibrium value of $x = -3.5$ could be used
for both populations although other values are not excluded.
Net emissivities integrated over the size distributions
are approximately equivalent to single-size particles with
radii of 135 $\mu$m and 15 $\mu$m for the large icy and
small silicate grains, respectively.

\subsection{Two Inner Warm Belts}

\subsubsection{Dust Properties and Locations}

{\it Spitzer} IRS, MIPS, and MIPS-SED data
reveal the presence of material within the
sub-mm ring's central void.
PSF analysis of 70 and 160 $\mu$m image surface brightness
profiles shows emission closer to the star than
11\asec~(35 AU) but also reveals that the emission
does not extend all the way to $r = 0$.
The local maximum and plateau in the IR excess
around 20--30 $\mu$m indicate emission at $T \sim 100-150$ K,
confirming the location of material at $r < 35$ AU, because at 35 AU,
$T_{dust}$ would be only $\sim$ 40 K for blackbody grains
or $\sim$ 60 K for 1-$\mu$m silicate grains.
Quadrature subtraction of a point source
size from the MIPS-SED 1-D source extent at 55 $\mu$m
(Figure 3a) yields an emission radius of
very roughly 8\asec~(25 AU).

Modeling the properties of the inner material involved
first noting that the marked change in SED slope at about 20 $\mu$m
(Figure 5) is easily explained as a silicate mineralogical
band (e.g.\, Chen et al.\ 2006).  Trial-and-error experiments with
the Heidelberg DDS code showed that the detectability of the band sets
a rough upper limit on grain radius $ a \leq 20/2\pi$ $\sim$ 3 $\mu$m.
In contrast, "generic" grains of that size without mineralogical attributes
produce an emission SED that is significantly broader than observed,
as do ice grains.

The 15-30 $\mu$m segment of the $\epsilon$ Eri SED can be matched
approximately by emission from a belt of warm ''astronomical''
silicate grains with characteristic size $\sim$ 3 $\mu$m,
located about 3 AU ($\sim 1$\asec) from the star.
This innermost warm belt is responsible for $\ga$ 95\%
of the system's excess emission at 25 $\mu$m first
detected by IRAS.  Note that dust located 3 AU from $\epsilon$ Eri
would be entirely contained within the IRS slits
that have a minimum width of 4.7\asec~(15 AU).

The precise position of the ''3 AU'' belt is constrained
by relations between grain size, radiative efficiency,
and temperature. For example, if the model grain size $a$ is
set at 2.2 $\mu$m instead of 3 $\mu$m, the belt must move to
$r \sim 3.8$ AU to have the correct color temperature.
However, grain efficiency declining steeply to longer
wavelengths for such small grains would
unacceptably ''erode'' the 20--30 $\mu$m SED plateau.
Thus, the grain size and belt location are
determined to perhaps 25\% precision.
There is no separate information regarding
an upper limit to the belt's width,
but that can be taken to be equivalent to the uncertainty
in belt location, i.e. $\sim$ 25\%.

The rising IRS spectrum near 35 $\mu$m (Figure 5)
cannot be fit by emission from the material inferred at $\sim$ 3 AU,
which has an SED decreasing beyond 30 $\mu$m.  The steep upward slope
from 35 to 60 $\mu$m, characterized by a color temperature of about 55 K,
plus the relatively narrow SED peak at 70--100 $\mu$m, locate the
next lower-temperature population of grains also
in a belt with narrow ranges of temperature and particle size.
The precise properties of this material are
especially constrained by the MIPS SED-mode 55--90 $\mu$m
flux densities, but because of slit losses, those data
must be renormalized using models of the source
geometry, as described in the next section.

A belt of $a \sim 8 \mu$m ''astronomical'' silicate grains
located at r $\sim 20$ AU ($\sim 6$\asec) satisfies
the SED and source size constraints.
Models placing the belt at 20 AU
can be distinguished from alternate locations at 10 or 30 AU.
Material at 20 AU could also explain the slight filling-in
of the first Airy dark ring at 24 $\mu$m, $r~\sim$ 23 AU
(Figure 6a).
Silicate grains of this size and temperature would not
show mineralogical features in the {\it Spitzer} data,
so the composition of this belt is consistent with,
but not determined to be, silicate.

Models with continuous distributions of material extending
across the sub-mm void produced too much total mid- and far-IR
flux and/or did not match the surface brightness profiles
and source size measurements. The warm material is clearly
restricted to annular zones with relatively empty gaps.

\subsubsection{Modeling Details, Including MIPS SED Normalization}

Convergence on final models for the two inner warm belts
involved iterative comparison of model SEDs from the
Heidelberg DDS software with the IRS 15-35 $\mu$m
and 55--90 $\mu$m (MIPS SED-mode) flux densities not already
accounted for by the model of the resolved disk.
Trial models yielded intrinsic surface brightness profiles
at the observed wavelengths that were:
(1) convolved with the 24 $\mu$m PSF to compare with the observed
unresolved source,
(2) convolved with a derived {\it Spitzer} 55 $\mu$m PSF
and MIPS SED-mode slit mask to compare with the observed
source 55 $\mu$m 1-D FWHM, and
(3) added to the already-established 70, 160 and 350 $\mu$m
intrinsic surface brightness profiles of the disk and then
PSF-convolved to check that the profiles were not significantly
perturbed by the extra warm material.

The model properties of these two belts within the sub-mm void
interact, in that they both produce significant flux density at
$\lambda$ $=$ 30--35 $\mu$m.  For example, changing the second
warm belt's properties toward smaller spatial radius and/or
particle radii produces greater 30--35 $\mu$m model emission,
requiring the innermost belt to have greater spatial radius
and/or particle radii so as to preserve the total mid-IR SED shape.

The slit losses and
flux normalization of the MIPS SED data were derived
by generating 2-D synthetic images of the source using model radial profiles
of the disk computed at the MIPS SED wavelengths. This analysis was restricted
to the two extremes of the MIPS SED wavelength range, 55 and 90
$\mu$m, using the midpoint at 70 $\mu$m as
the flux normalization reference.

Synthetic 2-D
images at 55, 70 and 90 $\mu$m were derived by convolving azimuthal
projections of the radial disk model profiles with the STinyTIM PSF, and then
adding a point source rescaled to the stellar flux at each wavelength.
The resulting 70 $\mu$m model ''image'' closely matches the
70 $\mu$m observations (Figure 6b).
A software mask with the size of the MIPS SED slit was then
centered on the synthetic
images to measured the fraction $q_\nu$ of the total flux recovered
through the SED-mode slit:  58\%, 47\% and 38\% respectively at
55, 70 and 90 $\mu$m.
The overall flux normalization factor $N^{flux}_{70}$ at 70
$\mu$m is derived by rescaling the SED in Figure 3b to have
a 70 $\mu$m flux density identical to the MIPS 70 $\mu$m image total
flux density (1.688 Jy, Table~1)
which yields $N^{flux}_{70} \simeq 1.78$. The
correction for the other two wavelengths can be derived
by factoring out the slit loss at 70 $\mu$m and multiplying
by the slit losses at 55 or 90 $\mu$m:

% equation 1
\begin{equation}
N^{flux}_\nu = N^{flux}_{70} \times \frac{q_{\nu}}{q_{70}}
\end{equation}

\noindent
This procedure produced corrected MIPS SED flux densities
of $1.62 \pm 0.25$ Jy at 55 $\mu$m and $1.64 \pm 0.25$ Jy at 90 $\mu$m.
Uncertainties were estimated by quadrature addition of the
70 $\mu$m uncertainty to the uncertainties in the uncorrected
flux densities plotted in Figure 3b.
Once the photospheric contribution is subtracted, an excess from the
disk of 1.30 $\pm$ 0.25 Jy at 55 $\mu$m and 1.52 $\pm$ 0.25 Jy
at 90 $\mu$m is obtained.

\subsection{Model Summary}

As described in the preceding section,
the over-all model of $\epsilon$ Eri's circumstellar material
includes:  (1) particles with low far-IR emissivity
and high sub-mm emissivity, consistent with the properties of
radius $a =$ 100--200 $\mu$m (effective $a \sim$ 135 $\mu$m)
amorphous H$_2$O ice grains in the sub-mm ring at $r =$ 35--90 AU,
(2) particles with high far-IR emissivity and low
sub-mm emissivity, consistent with the properties of
$a =$ 6--23 $\mu$m (effective $a \sim$ 15 $\mu$m)
''astronomical'' silicate grains at $r =$ 35--110 AU,
corresponding to the sub-mm ring plus an exterior halo,
(3) a narrow belt at $\sim$ 3 AU (T $\sim$ 120 K) of small
($a \sim$ 3 $\mu$m) silicate grains, and
(4) a narrow belt at $\sim$ 20 AU (T $\sim$ 55 K) of small grains
($a \sim$ 8 $\mu$m) of undetermined, but possibly silicate,
composition.
Specific properties of the model components are presented
in Table 2.

Figure 7 displays the SED of the complete model and the
separate disk components compared with the photometric and
spectrophotometric data (model photosphere SED subtracted).
{\it Spitzer} mid- and far-IR data especially
reveal a complicated SED shape that strongly constrains
the temperatures, locations, and grain sizes of warm unresolved
material.  The model is not unique but was built from the fewest
components with the simplest assumptions that produced
a good match to all the available data.
Many alternate models were tested, resulting in confidence that:
(a) 70 and 160 $\mu$m emission does not extend beyond 110 AU;
(b) 350 $\mu$m emission does not extend beyond 90 AU;
(c) 350 $\mu$m emission does not extend inside 35 AU;
(d) the dominant emission in the sub-mm ring is not
from silicate particles;
(e) 70 and 160 $\mu$m emission within 35 AU
must include non-emitting gaps, with emission
restricted to annular zones; and,
(f) grains in the innermost warm belt
at $r \sim 3$ AU have silicate composition.

The model rises less steeply than the IRS
observations at 10--18 $\mu$m.  The excess flux in
the model is less than 3\% of the emission from
the system in that range.  This may be related to uncertainty
in the photospheric model subtraction at short wavelengths
where the IR excess is barely significant, and may also
indicate the presence of crystalline silicates, with a
sharp 20 $\mu$m spectral feature, combined with
the amorphous silicates assumed in the model.

The top panel of Figure 8 displays the model radial profile
of perpendicular optical depth $\tau_\perp$, i.e.\ fractional
(unitless) surface density ($m^2$ of grain cross-sectional
area per $m^2$ of disk surface area; e.g.\ Backman 2004)
and in the bottom panel the corresponding mass surface density
inferred from the optical depth and model grain properties.
The slopes of the two functions across the sub-mm ring and halo
are not equal because the
model includes smooth variation in the particle size distribution
between the fiducial radii, so the ratio of particle cross section
(and thus optical depth) to mass surface density varies slightly
with radius.  The optical depths and mass densities of the two
inner rings in Figure 8 are calculated by assuming
$\Delta$$r$ values of 1 AU at 3 AU and 2 AU at 20 AU;
they would scale inversely with $\Delta$$r$.

The total inferred mass in {\it detected} grains,
i.e.\ with sizes
from a few microns to a few hundred microns, carrying most
of the radiating surface area, is roughly $5 \times 10^{-3}$
M$_\oplus$ $\sim$ 0.4 M$_{\rm Luna}$,
90\% of which is in the large (''icy'') grain
component of the sub-mm ring.
The total bolometric fractional luminosity ($L_d/L_*$)
of all the disk components is 1.1$\times$10$^{-4}$.

Figure 9 is a 2-D representation of the $\epsilon$ Eri
debris disk system in linear spatial scale, showing the
sub-mm ring, halo, and inner warm belts.
The position of the innermost warm belt is compared
in the inset figure with one orbit solution for the candidate
radial velocity planet (Section 5.4).

\section{Discussion}
\label{sec:discussion}

\subsection{Some Implications of the Disk Model}

The derived surface density profile for the
small grain component (Figure 8) has a nearly constant value
from 35 to 110 AU, consistent with dynamics controlled
by Poynting-Robertson radiation drag or its corpuscular
wind analog (Section 5.3).  In contrast, the large grain
component has surface density rising from 35 to 90 AU.
The larger grains are less subject to drag forces, so
their radial distribution may reflect
that of the underlying parent body population.

The void interior to the sub-mm ring is inferred to be partly
filled by two additional belts of particles with mean sizes
in both cases smaller than in the sub-mm ring.
The large grain population in the ring appears to be absent
in the void, even though 100 $\mu$m-size ice grains could
persist against sublimation for as long as $\epsilon$ Eri's age,
850 Myr, at distances as small as 3 AU from the star
(see equations 16 and 17 in Backman \& Paresce 1993
for expressions of ice sublimation time scales).
Moreover, the mid-IR SED, plus the 70 $\mu$m surface brightness
profile in particular, indicate that there are gaps in the distribution
of material from just beyond 3 AU to almost 20 AU,
and from just beyond 20 AU to 35 AU.
A mechanism is needed to prevent significant amounts
of both small and large grains from drifting via P-R drag from
the sub-mm ring into the inner void (Section 5.4).

The total mass of detected grains in $\epsilon$ Eri's innermost
belt is about 0.1$\times$ the mass estimated for the
warm belt detected around the 230 Myr-old A star
$\zeta$ Lep (Chen \& Jura 2001).
The size and structure of the $\epsilon$ Eri system, with
two warm belts in a central void surrounded by a cold disk,
and silicate dust identified in the innermost ring, bear
striking resemblance to the circumstellar material
in the much younger HD 113766 system (10-16 Myr, F3/F5 binary)
(Lisse et al.\ 2008).

An upper limit to the ratio of ice versus silicate grain mass
for the sub-mm ring region in which they co-exist is about 16.
This ratio is intermediate between values for
solar composition ($\sim$ 50) and for
comets ($\sim$ 2-5, e.g.\ Altenhoff et al.\ 2002).

\subsection{Comparison with Prior Studies of $\epsilon$ Eri}

\subsubsection{Sub-mm Ring Features}

As noted above, the sub-mm ring at 350 $\mu$m (Figure 4)
has the same inner and outer diameters observed in previously
published 850 and 450 $\mu$m maps (Greaves et al.\ 1998; 2005),
but 350 $\mu$m ring surface brightness features do not in
general align with 450 or 850 $\mu$m features.
The (JCMT / SCUBA) 450 and 850 $\mu$m data 
were obtained via dithered (''jiggle''-)
mapping, whereas the (CSO / SHARC II) 350 $\mu$m map was made
via continuous spiral scanning.  Real features should
be robust against such differences in data acquisition and
reduction procedures.

The strongest 450 and 850 $\mu$m feature, about 1.5$\times$
brighter than the ring average, is in the south-east sector.
A south-east feature appears in some form in all the maps
and was suggested to be a real part of the $\epsilon$ Eri
system by Greaves et al.\ (2005) based on its motion
over the 1997--2002 time span. Poulton, Greaves \& Cameron
(2006) made a careful statistical analysis of this issue.
Only a weak feature is seen at that location in the 350 $\mu$m
image.  Other fairly prominent 850 $\mu$m ring features,
especially in the south-west sector,
were considered by Greaves et al.\ to be background sources
although they have no clear counterparts at 450 $\mu$m.

The strongest 350 $\mu$m feature is
located in the north-north-east section of the ring, with
a surface brightness of about 0.24 mJy per square arcsec.
Its location corresponds approximately to a weak
450 $\mu$m feature and a moderately strong 850 $\mu$m
feature that could also be part of the $\epsilon$ Eri
system, considering proper motion plus orbital motion.
Recall (Section 3.1.3) that the 350 $\mu$m data,
when split into equal parts
and composed into separate maps, have about the same level
of disagreement regarding azimuthal features as
between maps at different wavelengths, and between
the 850 $\mu$m maps for 1997--2002 versus 2000--2002 
(Greaves et al.\ 2005).
As discussed in Section 3.1.3 and displayed in Figure 2,
no surface brightness features are seen in the ring at
70 and 160 $\mu$m.

Real density features need not have the same surface brightness
ratios versus wavelength as the main ring.  For example,
smaller grains trapped in planetary
resonances can have increased libration widths, smoother
distributions, and show less conspicuous clumping than
large grains, so that some features prominent at long
wavelengths may be less apparent at short wavelengths
(Wyatt 2006).  Azimuthal features with surface brightness
that does not vary progressively with wavelength, however,
are difficult to reconcile with physical models.

The disagreement between details of the different sub-mm maps
illustrates the difficulty of working at wavelengths with
large variable sky opacity and  emission.
These comparisons imply that the {\it Spitzer} far-IR images
and CSO 350 $\mu$m maps cannot confirm the ring
surface brightness features reported at 450 or 850 $\mu$m.

\subsubsection{Optical \& Near-IR Upper Limits}

The brightness of scattered light from circumstellar material can be
estimated from:

% equation 2
\begin{equation}
R_{\lambda, r} = (~\tau_\perp~/~4\pi~)~(~d^2_{\rm pc}~/~r^2_{\rm
 AU}~)~(~P~\phi(\gamma)~/~cos(i)~) 
\end{equation}

\noindent
in which R is the ratio, at a given wavelength and position,
of circumstellar surface brightness per square arcsec
to the star's brightness; $\tau_\perp$ is the
perpendicular optical depth of the circumstellar material;
$d_{\rm pc}$ is the system distance from the observer;
$r_{\rm AU}$ is the distance of the material
from the star; $P$ is the geometric albedo; $\phi(\gamma)$
is the phase function value for the star-dust-observer angle $\gamma$;
and $cos(i)$ is the inclination of the disk to the plane of the sky,
$i = 0$ representing face-on presentation.
A value of $cos(i)$ $\sim$ 1 can be used for the nearly face-on
$\epsilon$ Eri disk.
A value $\phi(\gamma)~\sim~0.5$ can be assumed
for particles much larger than the
wavelength of interest in an approximately face-on disk.

Proffitt et al.\ (2004) used the STIS spectrographic camera on HST
to search for scattered light from the sub-mm
ring region.  They set an upper limit of $m_{\rm STIS} \ga +25$ for
circumstellar light at $r \sim 55$ AU (17\asec) in a wide bandpass with
$\lambda_{eff} \sim 0.6$ $\mu$m.
Assuming an albedo of $P = 0.1$ for the model large grains,
which have the
highest optical depth in the sub-mm ring region (Figure 6a),
yields a  predicted scattered light surface
brightness at 55 AU almost exactly equal to the HST/STIS
limit.  Assuming instead an albedo of 0.9, possible
for fresh pure ice surfaces, would yield an optical surface
brightness almost 2.5 mags brighter than the limit.
In comparison, the model of the 2nd warm belt
($r \sim 20$ AU $\sim 6$\asec) predicts, an optical surface
brightness of $m_V \sim +21.7$ per square arcsec assuming
$P = 0.1$ and radial width $\Delta$$r = 2$ AU, brightness
scaling inversely with $\Delta$$r$.

Di Folco et al.\ (2007) report near--IR (K' band, 2.1 $\mu$m)
interferometric observations of $\epsilon$ Eri and $\tau$ Ceti 
with the CHARA array.  They resolved the stellar diameters in
both cases and detected circumstellar emission near $\tau$
Ceti but set an upper limit of 0.6\% of the photosphere for
emission within 1\asec~(3 AU) of $\epsilon$ Eri.  The present
model innermost belt predicts a total scattered light brightness
of approximately 3\% of the Di Folco et al.\ upper limit
if the grain near-IR albedo is assumed to be 0.1.  The direct
thermal emission at K' band from that belt, in fact from the
entire disk, would be many orders of magnitude fainter than
the scattered light brightness.

\subsubsection{SED Models}

Dent et al.\ (2000),
Li, Lunine \& Bendo (2003) (hereafter LLB03),
and Sheret, Dent \& Wyatt (2004) (hereafter SDW04)
combined pre-{\it Spitzer} broadband photometry and sub-mm map data to
produce models of the $\epsilon$ Eri disk with specific grain
spatial distributions, size distributions and compositions.
Given the limited data available to constrain them, those models are
consistent with the model presented in Section 4 of this paper
based on higher-precision observations over a much broader
wavelength range.

Dent et al.\ (2000) assumed non-porous
grains, with a simplified emissivity law determined
only by particle size without mineralogical features.
Their model requires material as close as
10 AU from the star with $\sim10$\% of the surface
density in the main ring.
LLB03 and SDW04 calculated models
assuming real grain mineralogies, including a range of porosities
(fraction of silicate grain volume occupied by vacuum or ice)
ranging from $p =$ 0 to 0.9.  SDW04 included their own
sub-mm photometric data regarding $\epsilon$ Eri
to supplement published maps.
LLB03's best-fit model involved highly porous
($p = 0.9$) grains, whereas SDW04's suite of
models favored non-porous (solid) grains.
The present model assumes non-porous grains
and agrees qualitatively with the SDW04 result
that the grain size distribution cannot extend down to
1 $\mu$m without producing too much short-$\lambda$ emission.
SDW04 remarked that the lack of $\mu$m-scale grains
is a puzzle given the inability of $\epsilon$ Eri's
radiation field to eject grains of that size.
Radiation pressure and corpuscular wind effects
are discussed further in Section 5.3.

The total circumstellar grain mass derived by LLB03 was
$\sim7 \times 10^{-3}$ M$_\oplus$ $\sim$ 0.6 M$_{Luna}$,
compared with 0.1 M$_{Luna}$ in SDW04's model and
0.4 M$_{Luna}$ in the current model.  The differences
between these mass estimates result from differing assumptions
about the particle properties and size ranges,
since each group is matching approximately the same
system fluxes, requiring similar total particle
radiating areas.

\subsection{Particle Dynamics}

\subsubsection{Grain Ejection via Wind versus Radiation Pressure}

Small grains are subject to radiation pressure (e.g.\ Burns et al.\
1979) and corpuscular winds (e.g.\ Plavchan et al.\ 2005)
that tend to remove them from their source
region, either rapidly outward by direct ejection for particles smaller
than the ''blowout'' size, or, for larger particles,  slowly inward
toward encounters with planets or sublimation near the star.
Grains also collide destructively.
Competition between rates of destructive processes versus rate of grain
resupply from larger parent bodies determines a
debris disk's equilbrium density and also whether there is net mass
flow outward or inward from the grain source (e.g.\ Wyatt 2005; Meyer
et al.\ 2007).  These processes will be considered in this and the
following subsection in view of the $\epsilon$ Eri system's
characteristics.  Porous and non-spherical grains would be affected
more strongly by radiation and wind than the spherical and non-porous
grains assumed here, so destruction and drift
time scales calculated below should be considered upper limits.

The direct (radial) radiation pressure (force per area) opposing
gravitational attraction between star and grain is:

% equation 3
\begin{equation}
P_r(radiation) = L_* / 4 \pi r^2 c
\end{equation}

\noindent
whereas the direct wind pressure is:

% equation 4
\begin{equation}
P_r(wind) = v_{wind} (dM_* / dt) / 4 \pi r^2 
\end{equation}

\noindent
The wind mass loss rate for $\epsilon$ Eri has been determined to be 
$\sim 3-4 \times 10^{10}$ kg s$^{-1}$
(Stevens 2005; Wood et al.\ 2005),
about 20--30$\times$ the solar value.
Assuming conservatively that young, chromospherically active
$\epsilon$ Eri has the same wind terminal speed of $\sim$ 400 km/sec
as the sun's, radial wind pressure is negligible
in the $\epsilon$ Eri system, less than 5\% as strong as
radiation pressure.  The star's mass/luminosity ratio is low
enough that solid silicate-density particles should be stable
against direct ejection down to sizes of a few $\times$ 0.1 $\mu$m.

This analysis indicates that:  (1) the model grain size lower
limits of a few microns in the sub-mm ring and inner unresolved
belts are not defined by radiation pressure or
neutral wind ''blowout'', and (2) radiation and wind
pressures in this system are too weak to move substantial
amounts of material outward, for example from the sub-mm ring
into the halo.

\subsubsection{Poynting-Robertson and Wind Azimuthal Drag versus Collisions}

The following two equations compare the azimuthal pressure
$P_{\theta}$ on orbiting grains respectively due to
radiation (P-R drag) and a neutral wind.

% equation 5
\begin{equation}
P_{\theta}(radiation) = v_{orb} L_* / 4 \pi r^2 c^2 
\end{equation}
%

% equation 6
\begin{equation}
P_{\theta}(wind) = v_{orb} (dM_* / dt) / 4 \pi r^2 
\end{equation}

\noindent
Using the values given above for wind mass and speed,
the radial drift rate for particles around $\epsilon$ Eri
is enhanced by a factor of about 20 relative to the purely
radiation-driven P-R drag rate (numerically similar,
by coincidence, to the $\epsilon$ Eri wind mass loss
rate relative to solar).

The timescale for grain-grain mutual collisions is:

% equation 7
\begin{equation}
t_{coll} \sim P_{orb} / 8 \tau_\perp
\end{equation}

\noindent
in which $P_{orb}$ is the orbit period and $\tau_\perp$ is
the perpendicular optical depth.
Cross-velocities as low as a few hundred meters per second
are sufficient to destroy small silicate and ice grains.
The solar system's asteroid and Kuiper belts both
have inclination and eccentricity ranges more than
large enough to result in routinely shattering
collisions, which may also be
assumed to pertain to the $\epsilon$ Eri disk.

At the inner edge of the sub-mm ring, $r = 35$ AU,
where P-R drag is most competitive with collisions,
the collision time scale for the large ($a \sim 135$ $\mu$m) grains
is $t_{coll} \sim 2 \times 10^{5}$ yrs, whereas the wind-drag time scale
for inward drift by $\Delta r = 1$ AU is $\sim 8 \times 10^5$ yrs.
Thus, even when enhanced by wind, virtually all the large icy grains
should be destroyed by collisions {\it in situ} and not be able
to drift inward to fill the central void.
For the small grains ($a \sim 15$ $\mu$m), on the other hand,
the collision destruction time scale
(including cross-sectional area of both small and large
grains as targets) is $\sim 1.5 \times 10^{5}$ yrs,
approximately the same as the wind-drag time scale
for these particles to drift 1 AU.
Given that the collision process is statistical, e.g.\
$\sim$ 1\% of grains can be expected to survive 5 $t_{coll}$,
some fraction of the small grain population {\it could}
travel inward from the sub-mm ring into the central void before
colliding with other grains.

At the inner and outer edges of the halo,
$r = 90$ and 110 AU respectively,
the collision time scale for the small grains is also
comparable to the $\Delta r = 1$ AU wind-drag drift time.
Thus, if grains are produced in the halo region they would
tend to stay in place or flow inward toward the sub-mm ring,
rather than outward.

\subsubsection{Collisional Scattering}

Is there a mechanism other than radiation or wind pressure that could
move material from the sub-mm ring to the outer halo, or does the
presence of significant material in the halo require a particle source
(presumably collisions of planetesimals) at $r \ga$ 90 AU?
In fact, scattering of material into new orbits after collisions
could suffice.

The {\it vis viva} equation expresses the
orbit properties resulting from velocity other than
circular orbit velocity at a given location,

% equation 8
\begin{equation}
v^2 = G M_* ( 2 / r  -  1 / a_{orb})
\end{equation}

\noindent
in which $v$ is the speed at radius $r$, and $a_{orb}$ is
the orbit semi-major axis.

Under what conditions would grain collisions in
the main, heavily populated ring be able to supply
material to the outer halo?
If grains or parent bodies collide at $r$
with average cross-speeds of $\xi \times v_{circ}$,
corresponding to relative orbit inclinations of order
$\delta$$i \sim \pm \xi$ rad and/or eccentricity ranges of
order $\delta$$e \sim \xi/2$, significant numbers of
fragments will leave the collision with
$\delta v \geq \xi$$v$ relative to the
motion of the colliding bodies' center-of-mass.
Collisions at $r \sim$ 90 AU with $\xi = 0.05$
would send some fragments into new orbits reaching
to 110 AU, the outer edge of the halo.
Therefore the extended halo observed at 70 and 160 $\mu$m
could be the result of collisions among planetesimals
and fragments moving with modest departure from
circular planar motion in the sub-mm ring.
If collsions are in fact the mechanism spreading material
from the ring into the halo, the lack of material
beyond 110 AU could indicate an upper limit
on inclinations and eccentricities in the ring.

\subsubsection{Particle Dynamics Summary}

Radiation plus wind pressure should not be able to eject
solid silicate-density particles larger than sub-micron size
from the $\epsilon$ Eri system, but the disk model has a
deficit of particles smaller than $a \sim 6~\mu$m in the
sub-mm ring.  Consideration of particle destruction time
scales suggests that a time span of order 10$^5$ years may have
elapsed since the last significant injection of small particles.
The small grain population has a spatial distribution indicating
overall control by azimuthal drag. Some small grains
near the inner edge of the sub-mm ring should be able
to drift into the central void faster than their collision
destruction time scale.  The fact that
this does not appear to be happening may indicate the presence
of a barrier planet preventing inward drift (Section 5.4).

The large grains in the sub-mm ring have collision time
scales much shorter than drift time scales, so they
are seen near their locations of creation, and their
radial mass density profile may track that of the
planetesimal parent bodies.

\subsection{Planets Around $\epsilon$ Eri}
\label{sec:infer_planets}

Radial velocity measurements of $\epsilon$ Eri
have suggested the presence of a planetary companion,
hereafter designated Planet A, with M~$\sin{i}=$ 0.86 M$_{Jup}$
and an orbit period of about 6.9 years (Hatzes et al.\ 2000
and references therein).  The corresponding orbital
semi-major axis is 3.4 AU, near the location inferred
for the innermost warm dust belt.

The eccentricity of 0.70 $\pm$ 0.04 reported for
the 3.4 AU planet by Benedict et al. (2006) corresponds
to periastron and apoastron distances of 1.0 and 5.8 AU, respectively.
A giant planet with this orbit would quickly clear that region
not only of dust particles but also the parent planetesimal belt
needed to resupply them, inconsistent with the
dust distribution implied by the {\it Spitzer} observations.
The disk's 70, 160 and 350 $\mu$m aspect ratios and position angles
reported here do not correspond to values of those quantities determined
from the 850 $\mu$m map, calling into question the
system inclination value used to combine the
astrometric and radial velocity data.
Butler et al.\ (2006) report a different orbital eccentricity
of 0.25 $\pm$ 0.23, which could
allow the planet to orbit entirely outside the innermost dust belt.

As discussed in Section 5.3, some of the small grains in the
sub-mm ring should be able to drift into the central
void and begin to fill it on timescales of $10^6$ to $10^7$ years.
A mechanism is needed to prevent grains from moving inward past
35 AU.  Liou \& Zook (1999) and Moro-Martin et al.\ (2005) modeled
the ability of massive planets to define the inner edges of
Kuiper belt-like dust rings. A hypothetical planet orbiting
at the edge of the main $\epsilon$ Eri sub-mm ring is here designated
Planet C.  The gravitational influence
of a planet near or in the ring has also
been proposed (Liou \& Zook 1999; Quillen \& Thorndike 2002;
Greaves et al.\ 2005) as the cause of ring density enhancements
corresponding to surface brightness features reported in
sub-mm imaging that may or may not be real (Section 5.2.1).

Confinement of a narrow debris belt at $r \sim 20$ AU
could indicate the presence of a third planet, here
designated Planet B.
Deller \& Maddison (2005) calculated dynamical models
of the $\epsilon$ Eri system including effects of planetary
masses, radiation pressure, and stellar wind, and
found that a planet orbiting at the inner edge of the sub-mm
ring would not by itself be able to define the ring's edge.
Rather, the effect of an additional Jovian-mass planet
with semi-major axis $a \simeq 20$ AU is needed.
Greaves et al.\ (2005) proposed that the mass
controlling some of the possible sub-mm ring features
may be located well inside the ring, at $r \sim$ 24 AU,
based on the orbital period suggested
by apparent motion of those features over 7 years.

Summarizing all the observations and models bearing
on the presence of planets around $\epsilon$ Eri, there may
be indications of three different planets in the system.
Planet A: the long-suspected but still unconfirmed
Jovian-mass planet with $a_{orb} = 3.4$ AU that may be associated
with the innermost warm debris belt detected by {\it Spitzer}.
Planet B: also perhaps Jovian-mass,
associated with the second warm debris belt at $r \sim 20$ AU
inferred in this paper, also possibly
helping define the sub-mm ring inner edge via 2:1 resonance,
keeping the zone between 20 and 35 AU clear.
(Note that a 1 $M_{Jup}$ planet with
age 0.85 Gyr located 6\asec~from $\epsilon$ Eri would not
have been seen by {\it Spitzer} but could be detected by JWST.)
Planet C: located at $r \sim 35$ AU, with mass less
than a few $\times$ 0.1 $M_{Jup}$ according to dynamical
models, preventing small grains from drifting inward
past the inner edge of the sub-mm ring.

\subsection{Simple Collisional-Evolutionary Model}

A dynamical model of the $\epsilon$ Eri sub-mm ring
was calculated assuming equilibrium between collisional
production of grains and removal by P-R plus corpuscular wind drag,
based on previous simple dynamical models of the solar system's Kuiper
belt (Backman, Dasgupta \& Stencel 1995) and solar-type circumstellar
debris disks observed by {\it Spitzer} (Meyer et al.\ 2007).
This model indicates that the
sub-mm ring can be maintained by collisions
of 11 M$_\oplus$ of parent bodies with assumed diameters of 10 km
and material densities of 1000 kg m$^{-3}$.  This agrees
with estimates made by Sheret, Dent \& Wyatt (2004) and
Greaves et al.\ (2005) using similar assumptions.  Thus, the
$\epsilon$ Eri system as observed can be in quasi-equilibrium.

{\it Spitzer} observations of Vega
revealed a far-infrared disk extending across the position of
a sub-mm ring into an outer halo (Su et al. 2005),
similar to the $\epsilon$ Eri system.  However, the Vega
disk's properties require recent injection of circumstellar material
to explain the abundance of short-lived particles.
The main difference between the two systems is that $\epsilon$ Eri's
luminosity (including wind)
is low relative to Vega's, so there is no
direct ejection of grains, P-R plus wind azimuthal drag rates are
lower, and equilibrium resupply by continuous minor collisions
can be effective.

A simple evolutionary model that estimates loss of belt
mass via destruction of planetesimal parent bodies, with subsequent
removal of fragments by the dynamical processes listed above,
was described and applied to the solar system's Kuiper belt,
asteroid belt, and extrasolar analogs by Meyer et al.\ (2007).
The same model, tuned to the properties of the $\epsilon$
Eri system, predicts that for an initial mass of planetesimal
parent bodies in the range 15--30 M$_\oplus$, similar to the
inferred original mass of the Kuiper belt (e.g.\ Stern \& Colwell
1997), about 11 M$_\oplus$ would remain at a system age of 0.85 Gyr.
In other words, $\epsilon$ Eri's current debris disk is similar
in total mass to what the solar system's Kuiper belt
mass would have been at the age of $\epsilon$ Eri had there
not been a prior Late Heavy Bombardment (LHB) clearing event
(e.g.\ Levison et al.\ 2007).

Collisional evolution is dissipative, eventually converging
on a planetesimal population that has a time scale for further
collisions proportional to the system age, nearly independent of the
initial conditions, if there are no external effects.
However, to transform the young solar system's
Kuiper belt or $\epsilon$ Eri's
debris disk into a low-mass system like the
present-day Kuiper belt, with at most a few
$\times$ 0.1 M$_\oplus$ at age 4.5 Gyr (Luu \& Jewitt 2002),
requires additional mass reduction processes
beyond quasi-steady state collisional grinding and debris removal
(Meyer et al.\ 2007 and references therein).
For example, in our solar system, the LHB event,
possibly driven by a late episode of planet migration, may have
destroyed most of the Kuiper belt's mass relatively sudddenly,
and similar events may happen in other systems with ages
of a Gyr or more (Levison et al.\ 2007 and references therein).
Alternatively, a process that is continually at work in our solar
system, planetary pertubations pulling planetesimals out of the
Kuiper belt, could, with the help of collisional grinding,
transform $\epsilon$ Eri's 11 M$_\oplus$ disk at $t =$ 0.85 Gyr
into the solar system's 0.1 M$_\oplus$ Kuiper belt at 4.5 Gyr
if the perturbative removal process
supplementing collisions were to operate with an
exponential time scale of about 1.5 Gyr.
In contrast, $\tau$ Ceti's debris disk has an inferred
mass of 1.2 M$_\oplus$ at an age of 8 Gyr, so it may
not have had a major disruptive episode (Greaves et al.\ 2004).

\section{Summary}

The planetary debris disk around $\epsilon$ Eridani
was observed using {\it Spitzer}'s IRAC and MIPS cameras,
{\it Spitzer}'s IRS and MIPS SED-mode spectrophotometers,
and the SHARC II bolometer array on the CSO sub-millimeter telescope.

The slight 25 $\mu$m excess detected by IRAS is confirmed
photometrically by the {\it Spitzer} IRS and MIPS data
but not resolved in the 24 $\mu$m image.
The CSO 350 $\mu$m map confirms the sub-mm ring
at $r \sim$ 11--28\asec~(35--90 AU),
previously observed at 450 and 850 $\mu$m,
about 2$\times$ the width of our solar system's Kuiper belt.
Analyses of the 70 and 160 $\mu$m images reveal emission
extending from within a few arcsec of the star outward
to $r \sim$ 34\asec~(110 AU),
both inside and outside the sub-mm ring.
The {\it Spitzer} 70 and 160 $\mu$m images do not have enhanced
surface brightness at the position of the ring.
The 350 $\mu$m surface brightness varies
noticeably with azimuth around the ring but the
locations of the brightest features do not correspond
between 350 $\mu$m data sets reduced separately,
or with with the brightest features in the longer-wavelength maps,
calling into question the reality of the sub-mm ring azimuthal
asymmetries.

The over-all system spectral energy distribution is complex,
with significant excess relative to the
photosphere beginning at $\lambda \sim 15~\mu$m.
The excess rises rapidly
to 20 $\mu$m, flattens from 20 to 30 $\mu$m, rises rapidly again
from 30 $\mu$m to a peak between 70 and 100 $\mu$m, and then
decreases toward longer wavelengths.

A model of the disk,
constrained by all the available data,
combines two populations of material in the resolved disk:
(1) large (radii $ a \sim$ 100--200 $\mu$m) solid grains
in the sub-mm ring, which have high emissivity at sub-mm wavelengths
but low emissivity at 70 and 160 $\mu$m, consistent with H$_2$O ice
composition, plus (2) smaller ($ a \sim$ 6--23 $\mu$m) solid grains in
both the sub-mm ring and its exterior ''halo''
(r = 25--34\asec, 90--110 AU) which have high emissivity
in the far-IR and low emissivity at sub-mm wavelengths,
consistent with silicate composition.
The mid- and far-IR spectrophotometry and radial source profiles
imply two additional narrow belts of material,
one at r $\sim$ 3 AU (1\asec) with a weak
20 $\mu$m silicate emission feature,
and the other at r $\sim$ 20 AU (6\asec)
of undetermined, but possibly silicate, composition.

The grains in the
halo may be spread there from the ring region via
the effect of collisions.  Definition
of the inner edge of the sub-mm ring, and confinement of the
two rings of material closer to the star, probably require
gravitational influences of three planets.
A dynamical model of the disk indicates that the observed
populations of grains in the sub-mm ring and outer halo can
be supplied in equilibrium by
collisions of about 11 M$_\oplus$ of km-scale parent
bodies located in the sub-mm ring.

\acknowledgements

This paper is based in part on observations made
with the {\it Spitzer Space Telescope} that is operated by the Jet 
Propulsion Laboratory, California Institute of Technology, under
NASA contract 1407, and in part on observations from the
Caltech Submillimeter Observatory (CSO).
Support for this work was provided by NASA through awards
issued by JPL/Caltech, including subcontracts 1278243
and 1255094 respectively to the SETI Institute and the
University of Arizona, and also by NASA grant NNG05GI81G.
Analyses in this paper made use of the SIMBAD
database operated at CDS, Strasbourg France.

We are grateful to Jane Morrison and Myra Blaylock, University of Arizona,
for help with MIPS image data processing, and Paul Smith,
University of Arizona, for guidance in MIPS SED-mode data processing.
We thank Peter Plavchan and Eric Becklin for helpful discussions.

%%
%% Tables
%%
%
\clearpage 

% Table 1
\begin{deluxetable}{rrrrl}
\tablecaption{$\epsilon$~Eri Broadband Photometry}
\tablehead{
\colhead{$\lambda$ [$\mu$m]} &
\colhead{Total $F_\nu$ [Jy]} &
\colhead{Star $F_\nu$ [Jy]} &
\colhead{Excess $F_\nu$ [Jy]} &
\colhead{Notes}}
\startdata
   3.550 & 64.320 $\pm$ 1.140 & 64.320 &   ---  & IRAC\\
   4.493 & 39.800 $\pm$ 0.700 & 39.800 &   ---  & IRAC\\
   5.731 & 26.110 $\pm$ 0.470 & 26.110 &   ---  & IRAC\\
   7.872 & 14.910 $\pm$ 0.260 & 14.910 &   ---  & IRAC\\
  12.000 &  6.690 $\pm$ 0.540 &  6.570 &  0.130 & Color-corrected IRAS\\
  23.680 &  2.042 $\pm$ 0.040 &  1.726 &  0.315 & MIPS\\
  25.000 &  1.900 $\pm$ 0.280 &  1.550 &  0.350 & Color-corrected IRAS\\
  60.000 &  1.750 $\pm$ 0.250 &  0.270 &  1.480 & Color-corrected IRAS\\
  71.420 &  1.688 $\pm$ 0.207 &  0.189 &  1.499 & MIPS\\
 100.000 &  1.990 $\pm$ 0.300 &  0.100 &  1.890 & Color-corrected IRAS\\
 156.000 &  0.970 $\pm$ 0.186 &  0.039 &  0.931 & MIPS\\
 200.000 &  1.880 $\pm$ 0.280 &  0.020 &  1.860 & ISO$^1$\\
 350.000 &  0.366 $\pm$ 0.050 &  0.008 &  0.358 & CSO / SHARC II\\
 450.000 &  0.250 $\pm$ 0.020 &  0.005 &  0.245 & JCMT / SCUBA$^2$\\
 850.000 &  0.037 $\pm$ 0.003 &  0.001 &  0.036 & JCMT / SCUBA$^2$\\
\enddata
\tablerefs{(1) Walker \& Heinrichsen (2000); (2) Greaves et al. (2005)}
\end{deluxetable}

\clearpage

% Table 2
\begin{deluxetable}{lcccccc}
\tablecaption{Model Components}
\tablehead{
\colhead{component} &
\colhead{$r$ [AU]} &
\colhead{$M_{\rm T}$ [$M_\oplus$]} &
\colhead{$\alpha$} &
\colhead{$a$ [$\mu$m]} &
\colhead{$x$} &
\colhead{$f$}}
\startdata
W1 & 3 & $1.8 \times 10^{-7}$ & -- & 3.0 & -- & $3.3 \times 10^{-5}$ \\
W2 & 20 & $2.0 \times 10^{-5}$ & -- & 8.0 & -- & $3.4 \times 10^{-5}$ \\
RS & 35--90 & $2.0 \times 10^{-4}$ & +0.01 & 6.0--23  & -3.5 & $3.0 \times 10^{-5}$ \\
RL & 35--90 & $4.2 \times 10^{-3}$ & +1.05 & 100--200  & -3.5 & $4.4 \times 10^{-6}$ \\
HS & 90--110 & $2.5 \times 10^{-4}$ & +0.15 & 15-23 & -3.5 & $4.8 \times 10^{-6}$ \\
\enddata
\tablecomments{Columns:  (1) model component: 
W1 = warm belt 1, W2 = warm belt 2, 
RS = sub-mm ring, small grains; RL = sub-mm ring, large grains;
HS = halo, small grains;  (2) location; (3) total mass;
(4) mass surface density exponent, assumed to be zero for the W1 and
W2 components, fitted to data for the other components;  (5) grain radius;
(6) assumed grain size distribution exponent;
(7) fractional luminosity, $L_d/L_*$}
\end{deluxetable}

%%
%% Figures and Captions.
%%

%
% 6-PANEL MIPS IMAGES
\clearpage

%\markright{Figure 1}
\onecolumn
\begin{figure}[t]
\begin{center}
\includegraphics[width=0.85\textwidth]{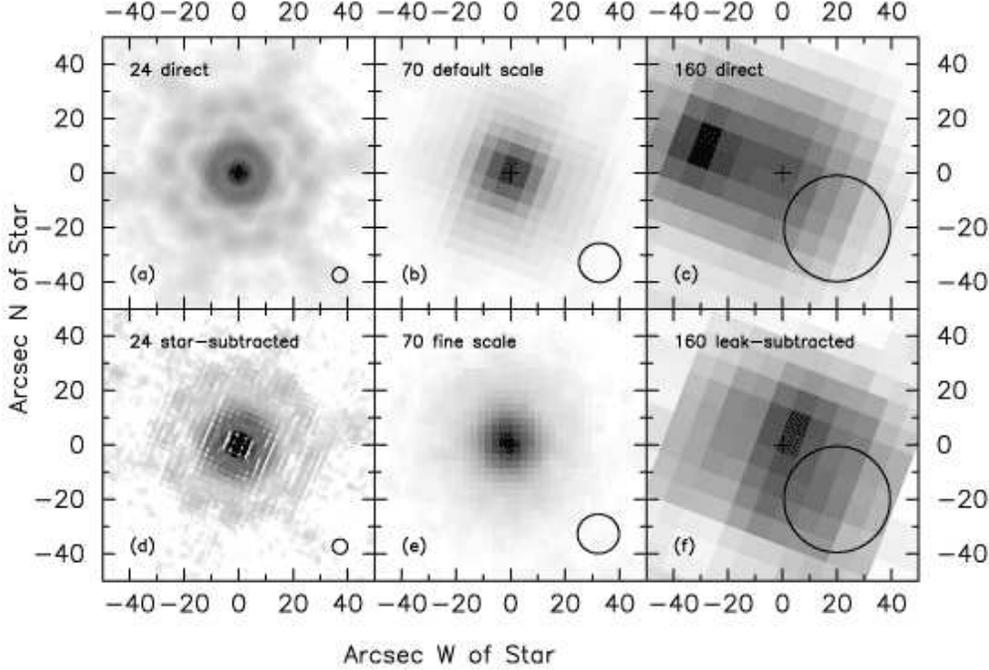}
\end{center}
\figcaption{Images of $\epsilon$ Eri in the three {\it Spitzer} MIPS channels.
N is up and E to the left in all the panels; 10\asec~$=$ 32 AU at
$\epsilon$ Eri's distance.
A cross marks the stellar position at the epoch of each observation,
taking into account $\epsilon$ Eri's proper motion of $\sim$ 1\asec~per year.
Dark is positive.  Circles indicate FWHM beam sizes.
{\bf The left column} shows 24 $\mu$m images 
in log stretch.  These have been reconstructed from multiple dithered 
exposures and are oversampled by a factor of four.  Panel $(a)$ is the 
direct image, dominated by the stellar PSF.  Panel $(d)$ shows this image 
after subtracting a PSF reference star scaled to the photospheric
flux density.  The remaining excess (18\% of the total 24 $\mu$m
emission) appears largely as an unresolved point source.
{\bf The middle column} shows 70 $\mu$m images in linear stretch.
Panel $(b)$
is the default pixel-scale image used for photometric measurements,
and panel $(e)$ shows the fine pixel-scale image.
{\bf The right column} shows 160 $\mu$m images in linear stretch.
Panel $(c)$
is the direct image, and panel $(f)$
shows this image after subtraction of the
ghost image arising from the spectral leak.
}
\label{mips-data}
\end{figure}
%

%
% 70 AND 160 MICRON DATA MINUS AZIMUTHALLLY SMOOTHED
\clearpage

%%\markright{Figure 2}
\onecolumn
\begin{figure}[t]
\begin{center}
\includegraphics[width=0.45\textwidth]{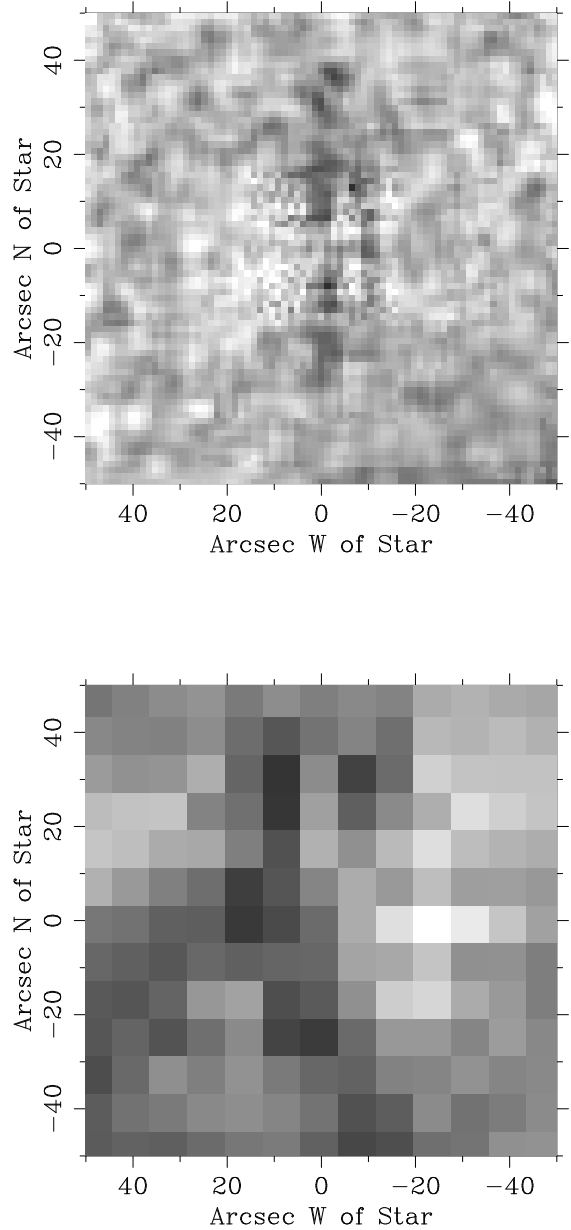}
\end{center}
\figcaption{
(a) difference between the {\it Spitzer}
MIPS fine-scale 70 $\mu$m image and
its azimuthal average.
(b) the same for the MIPS 160~$\mu$m image.
The vertical bars are due to
incomplete correction of column artifacts.
The remaining residuals are consistent with noise
from the subtraction, indicating absence of surface brightness
features in the sub-mm ring within the sensitivity limits.
}
\label{div_bumps_70f}
\end{figure}
%

%
% MIPS SED-MODE FLUX AND FWHM
\clearpage

%\markright{Figure 3}
\begin{figure}[t]
\begin{center}
\includegraphics[width=0.65\textwidth]{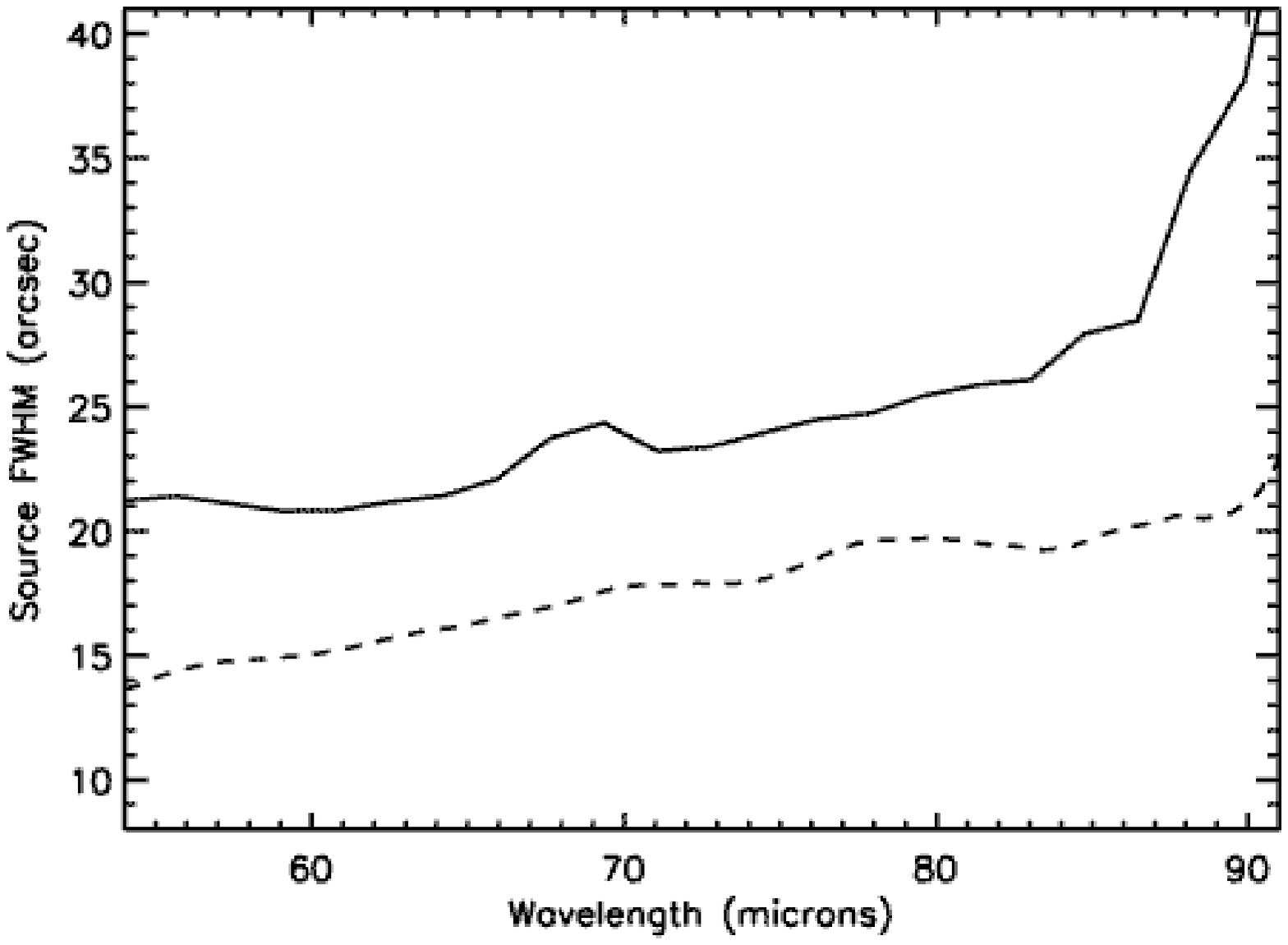}
\includegraphics[width=0.65\textwidth]{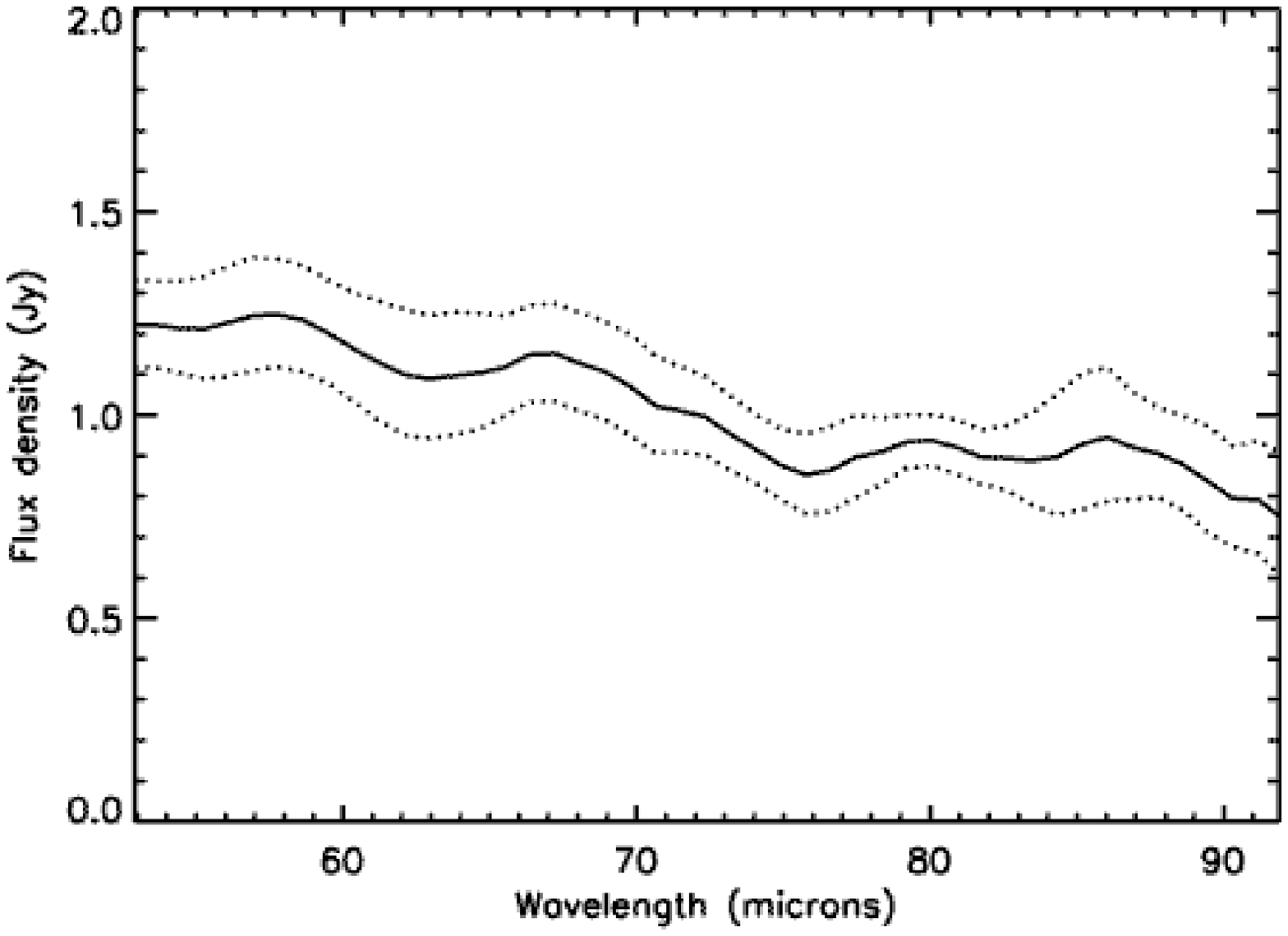}
\end{center}
\figcaption{
MIPS SED-mode measurements of $\epsilon$ Eri.
(a) (solid line) FWHM of the
$\epsilon$ Eri disk emission versus wavelength, extracted by Gaussian
fitting along the slit at PA 85\deg, versus (dashed line) the same for
Canopus, an unresolved
calibration star with no IR excess.  $\epsilon$ Eri is
clearly resolved across the wavelength range of the MIPS SED
measurements.
(b) Star+disk spectrum windowed by the 20\asec-wide slit,
without aperture correction.
The dotted lines indicated the range of photometric
uncertainty based on uncertainty in Gaussian fits
across the slit.  Subsequent model-dependent flux normalization
of these data is discussed in Section 4.4.2.
}
\label{mips-sed}
\end{figure}
%

%
% SHARCII CSO 350 MICRON IMAGE
\clearpage

%\markright{Figure 4}
\onecolumn
\begin{figure}[t]
\begin{center}
\includegraphics[width=0.95\textwidth,angle=-90]{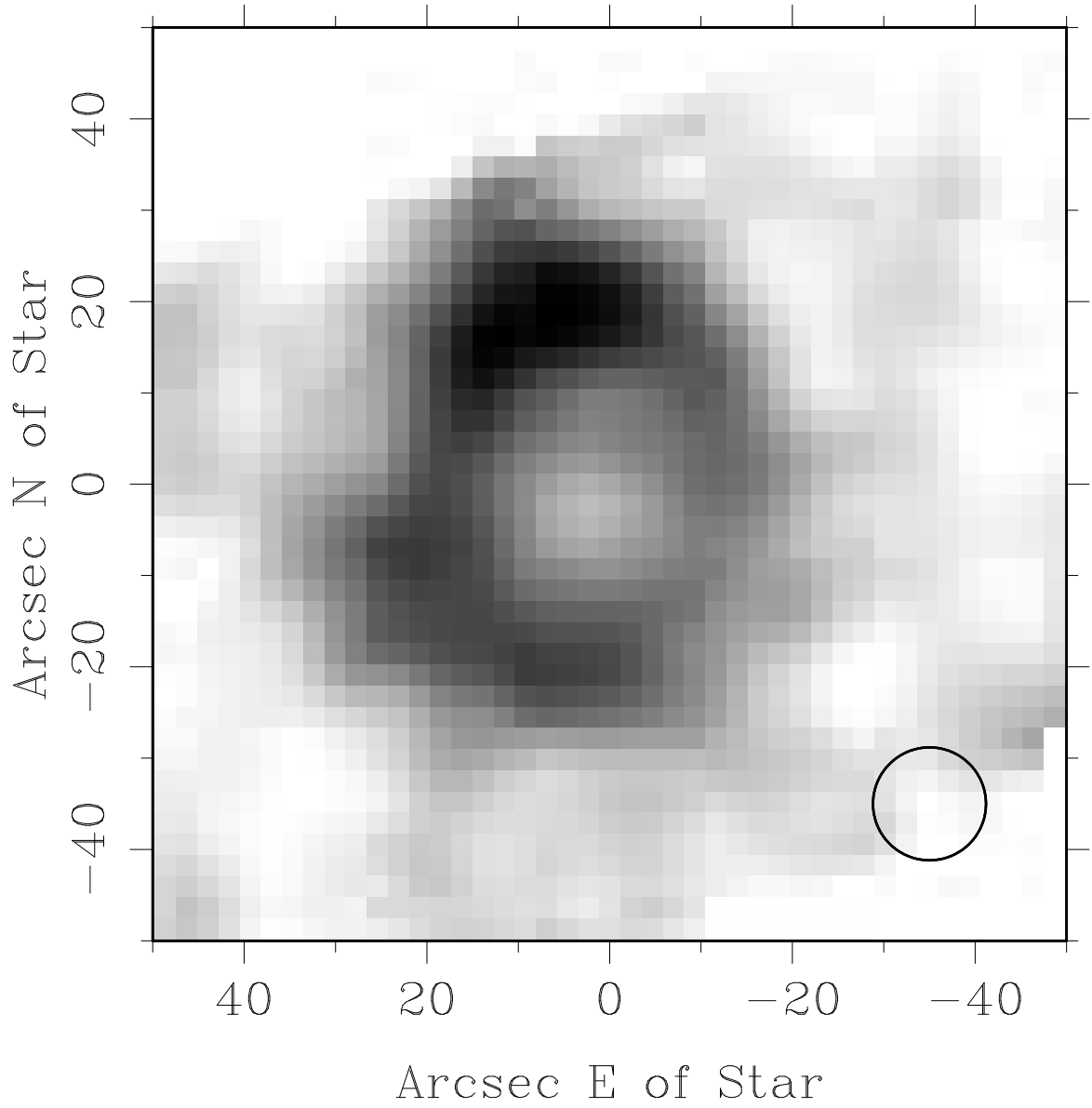}
\end{center}
\figcaption{
Caltech Submillimeter Observatory (CSO) SHARC II 350 $\mu$m map.
Dark is positive.
The circle's diameter indicates 12.7\asec~resolution to which the
map was smoothed, equivalent
to 41 AU at the distance of $\epsilon$ Eri.
A scaled PSF representing the relatively small 350 $\mu$m
brightness of the central star has been subtracted.
}
\label{350um-CSO}
\end{figure}
%

%
% MERGED BROADBAND SED (EXCLUDING MIPS SED)
\clearpage

%\markright{Figure 5}
\onecolumn
\begin{figure}[t]
\begin{center}
\includegraphics[width=0.65\textwidth,angle=-90]{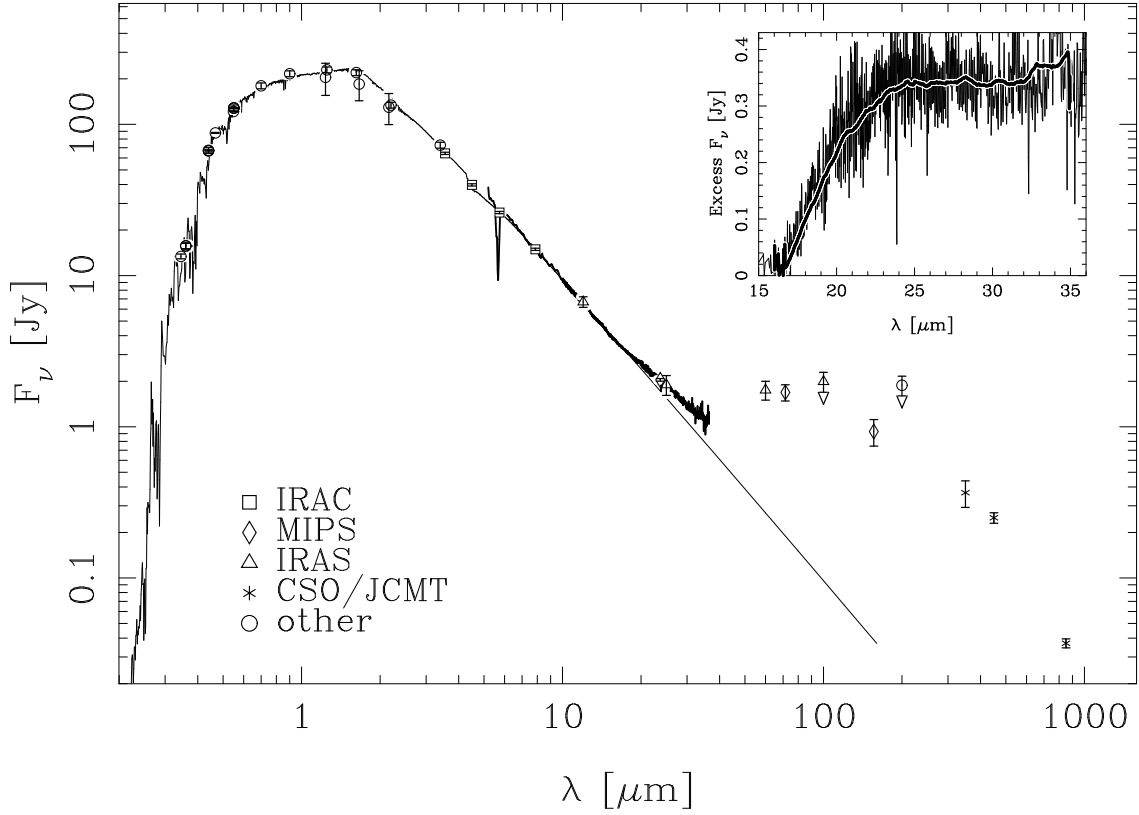}
\end{center}
\figcaption{
Merged broadband SED of the $\epsilon$ Eri system.  Numerical values are
presented in Table 1.  The thin solid curve is a Kurucz photosphere for
T = 5000 K, log g = 4.5, fit to the four {\it Spitzer} IRAC photometric
points at 3.6 to 7.9 $\mu$m.
The thick curve is the combined {\it Spitzer} IRS SL + SH + LH
spectrum rescaled to have zero average excess at 5--12~$\mu$m. The
IRAS 100~$\mu$m point and the ISO 200~$\mu$m point are
considered upper limits due to background contamination. 
The inset panel shows the mid-infrared segment of the SED from
{\it Spitzer} IRS (thin solid line), after subtraction of the
photosphere model displayed in the main figure. The thick line
is the same spectrum smoothed with a
111-point ($\sim 3.3$ $\mu$m) boxcar.
}
\label{merged-SED}
\end{figure}
%

%
% RADIAL PROFILES
\clearpage

%\markright{Figure 6}
\onecolumn
\begin{figure}[t]
\begin{center}
\includegraphics[width=0.65\textwidth,angle=-90]{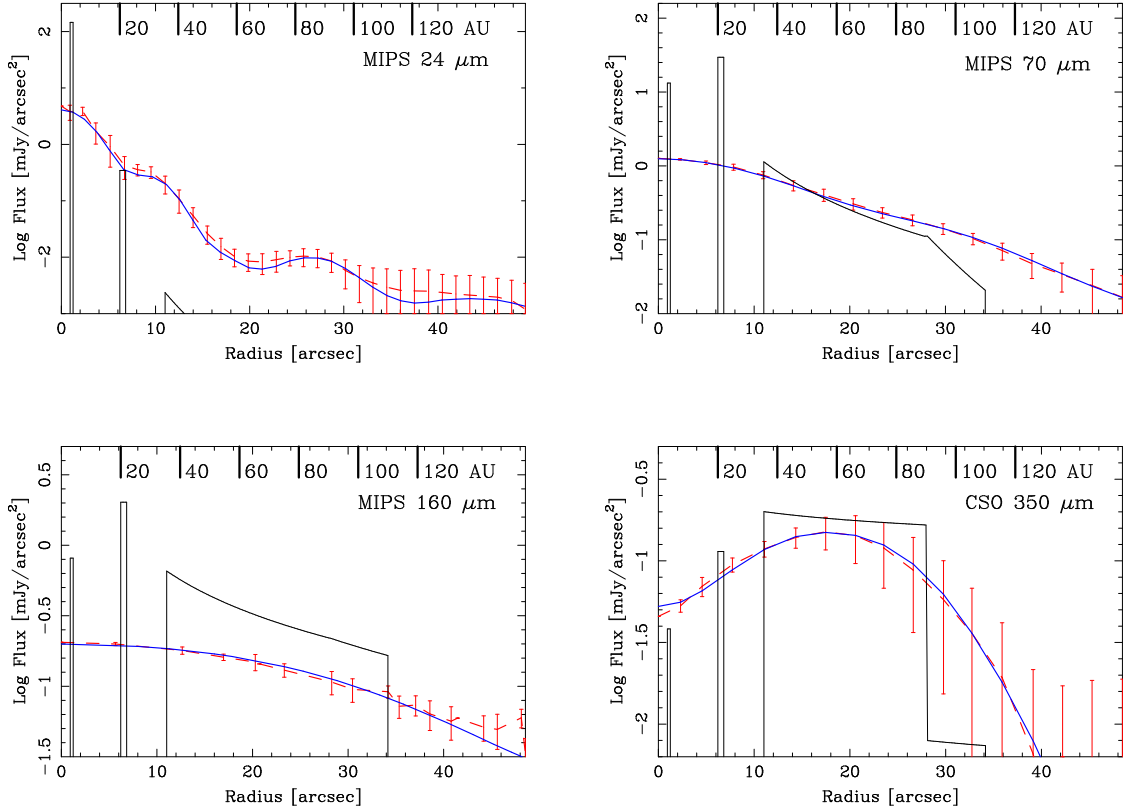}
\end{center}
\figcaption{
Radial profiles of images at 24, 70, 160 $\mu$m
({\it Spitzer} MIPS) and the 350 $\mu$m (CSO) map.
In each wavelength panel the dashed line with error bars represents
the azimuthal average of the image data after subtraction
of the stellar photosphere PSF. Observational error bars are
estimated as the rms of the azimuthal average.
The thick solid line is the disk model after convolution with
the relevant instrumental PSF, fitted to the observations.
The discontinuous thin solid line is the model
intrinsic surface brightness before PSF convolution (Section 4.2).
}
\label{1d_profiles}
\end{figure}
%

% 
% OVER-ALL SYSTEM SED AND MODEL
\clearpage

%\markright{Figure 7}
\onecolumn
\begin{figure}[t]
\begin{center}
\includegraphics[width=0.65\textwidth,angle=-90]{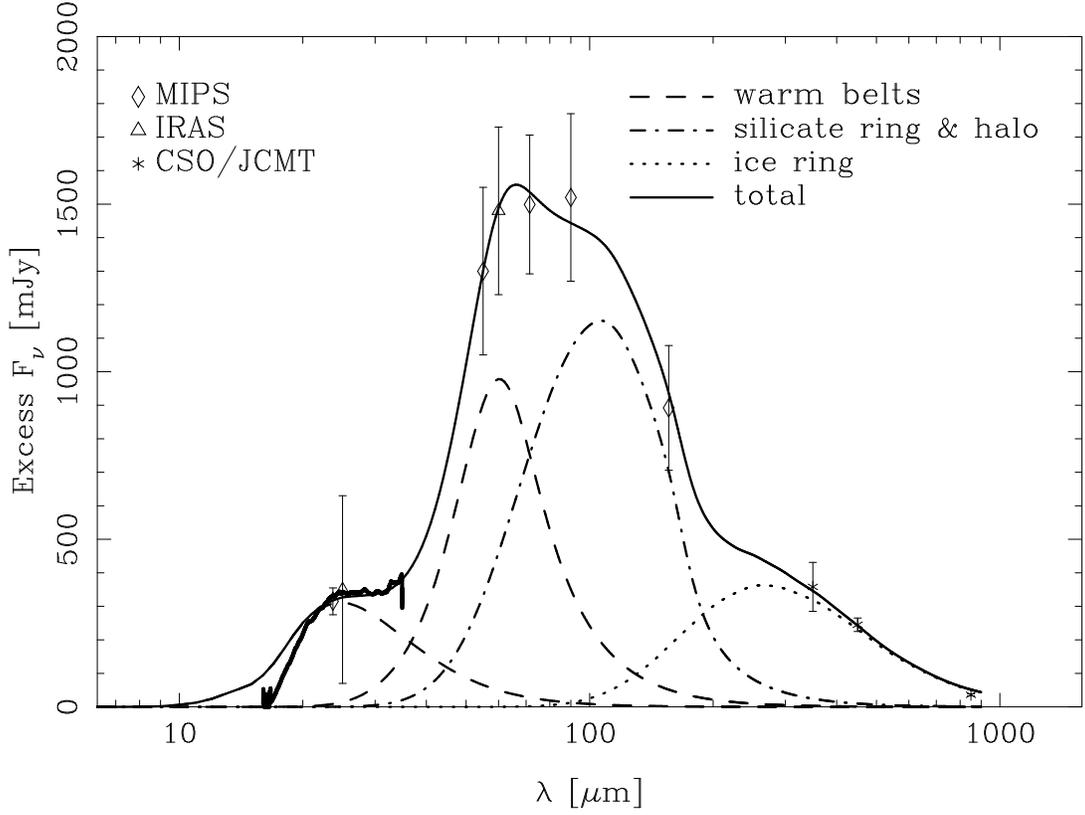}
\end{center}
\figcaption{
Observed SED of the $\epsilon$ Eri disk, after subtraction of
the stellar photosphere, compared with the model SED.
Individual photometric points include MIPS SED
flux densities at 55 and 90~$\mu$m, iteratively
aperture-corrected using model images, and
rescaled according to the measured MIPS 70 $\mu$m image total
flux. The thick solid line is the IRS SL + SH + LH combined spectrum,
rescaled to have zero average excess for $\lambda = 5-12 \mu$m.
The thin solid line is the total model flux (sum of all four dust
components). The two dashed lines are the contributions of the two
unresolved inner belts. The
dot-dashed line is the contribution of the
small ($a \sim 15~\mu$m) silicate grains in the sub-mm ring and halo.
The dotted line is the contribution of the
large ($a \sim 135~\mu$m) ice grains in the sub-mm ring.
}
\label{irs+phot}
\end{figure}
%

%
% NORMAL OPTICAL DEPTH & MASS SURFACE DENSITY
\clearpage

%\markright{Figure 8}
\onecolumn
\begin{figure}[t]
\begin{center}
\includegraphics[width=0.65\textwidth]{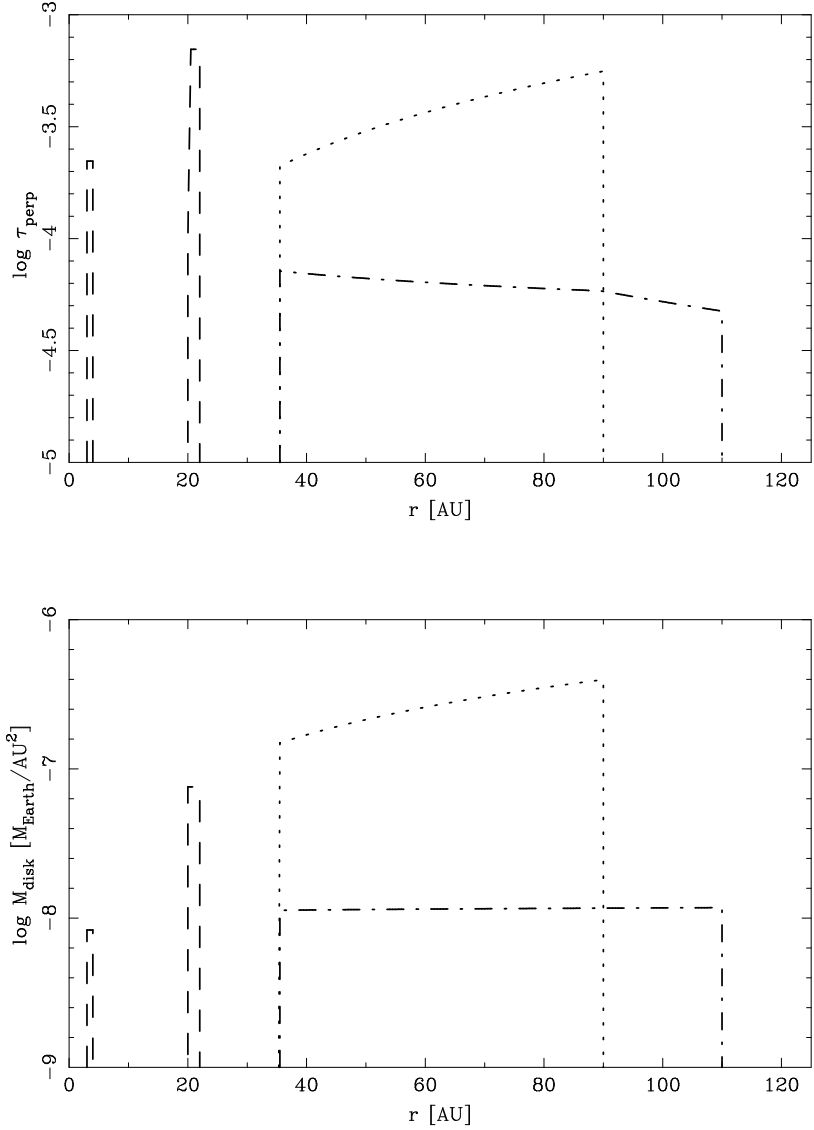}
\end{center}
\figcaption{
(a) Perpendicular optical depth $\tau_\perp$ versus
position for the five model disk components.
The two unresolved inner belts (dashed lines),
outer ''icy'' ring (dotted line), and
ring+halo ''silicate'' dust populations
(dot-dashed line) are shown.
(b) Mass surface density versus radius corresponding
to the optical depth profiles in panel (a), assuming
grain compositions
and material densities described in the text.
The difference in slopes between the optical depth
and mass surface density profiles is due to
radial variation in model particle properties.}
\label{tauprofile}
\end{figure}
%

%
% SYSTEM CARTOON
\clearpage

%\markright{Figure 9}
\onecolumn
\begin{figure}[t]
\begin{center}
\includegraphics[width=0.85\textwidth,angle=-90]{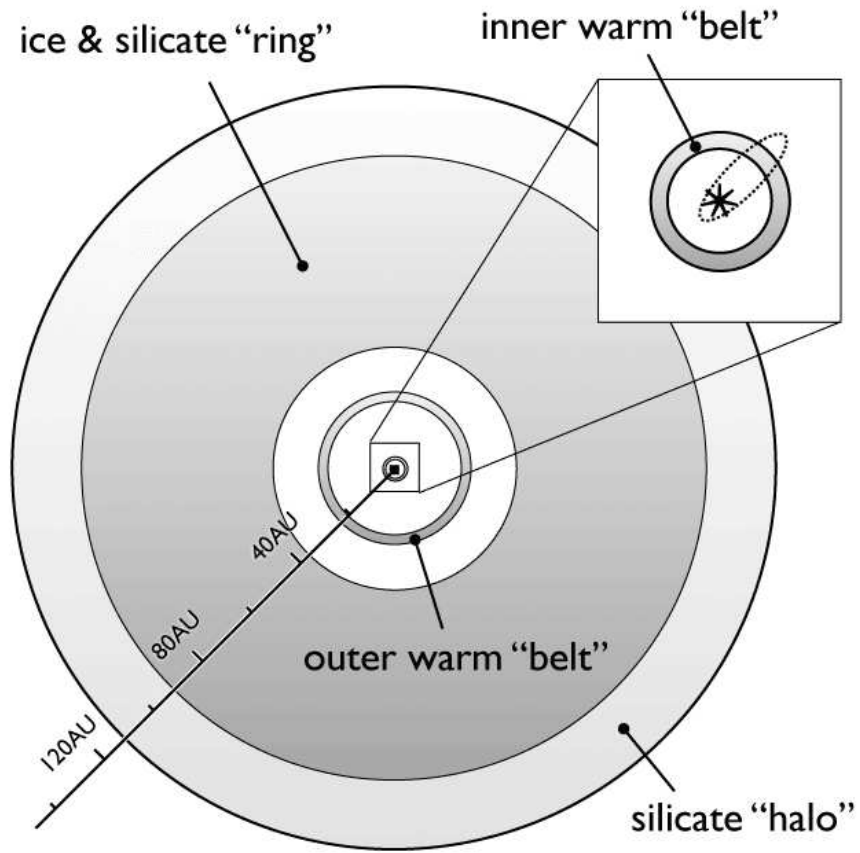}
\end{center}
\figcaption{
Representation of the $\epsilon$ Eri debris disk model components.
The small-scale dotted ellipse is one solution for the orbit of the
suggested radial velocity planet that appears to be inconsistent with
the innermost warm debris belt's position.
}
\label{cartoon}
\end{figure}

\end{document}